\documentclass[12pt]{article}

\usepackage{amsmath, amsfonts, amssymb, mathrsfs, graphics}
\numberwithin{equation}{subsection}

\usepackage{amsmath,amsfonts,amssymb}
\usepackage{mcite}
\usepackage{float}
\usepackage[usenames,dvipsnames]{xcolor}

\usepackage{graphicx}
\usepackage{hyperref}

\newcommand{\mathsym}[1]{{}}
\def\id{\protect{{1 \kern-.28em {\rm l}}}}

\def\be{\begin{equation}}
\def\ee{\end{equation}}
\def\ba{\begin{eqnarray}}
\def\ea{\end{eqnarray}}

\makeatletter
\renewcommand\section{\@startsection {section}{1}{\z@}%
                                   {-3.5ex \@plus -1ex \@minus -.2ex}%
                                   {2.3ex \@plus.2ex}%
                                   {\normalfont\large\bfseries}}
\renewcommand\subsection{\@startsection{subsection}{2}{\z@}%
                                   {-3.25ex\@plus -1ex \@minus -.2ex}%
                                   {1.5ex \@plus .2ex}%
                                   {\normalfont\normalsize\bfseries}}

\makeatother

\def\eps{{\varepsilon}}

\def \foot {\footnote}

\def \ha {{1 \over 2}}
\def \td {\tilde}
\def \ci{\cite}

\def\S{{\mathcal S} }

 \def \J {{\mathcal  J}}

\def\a{\alpha}

\def \a {\alpha}

\def\g{\gamma}
\def\ov{\over}

\def\J{{\mathcal J}}

\def\l{\lambda}

\def\foot{\footnote}

\def \ci {\cite}

\def \fo { {1\ov 4}}

\def \l  {\lambda}

\def \S {{\rm S}}

\def \td {\tilde}

\def \O {{\mathcal O}}

\def \l {\lambda}
\def\foot{\footnote}

\def \adss {$AdS_5 \times S^5~$ }

\def \ov {\over}

\def\cc{\circ}

\def \ha{{1\ov 2}}

\def \J {\mathcal{J}}

\def \S {{\cal S}}
\def \J {{\cal J}}

 \def \bb {\bar \beta}

\def\foot{\footnote}

\def \sql {{\sqrt \lambda}}

\def \cS {{\cal S}}

\def \ads {$AdS_5 \times S^5$\ }

\def \ov {\over}

\def \varpi {{\rm w}}
\def \OO {{\cal O}}

\def \te {\theta}

\def \S  {{\rm S}}

\def \C {{\cal C}}

 \def \J {{\cal J}}
 \def \S {{\cal S}}

\def \os  {\OO({\textstyle{ 1\ov \sqrt{\lambda}}})}

 \def \sql {\sqrt{\lambda}}

\def \cc {{c }} 
\def \OO {{\cal O}}
\def \te {\textstyle}
\def \fl {\sqrt[4]{\lambda}}

\def \fo {{\textstyle{1 \ov4}}}
\def \rx {{\rm x}}
\def \hg {{\hat \g}}

\def \C  {{\rm C}}
\def \hC  {{\rm \hat  C}}
\def \dd  {{\rm d}}
\def \bb {{\rm b}}
\def \dDelta {2}
\def \sql {{\sqrt{\lambda}}}
\def \ed {\end{document}}
 \def \an {{\rm an}} \def \nan {{\rm nan}}

\def \td {\tilde}
 
\newcommand{\mc}{\mathcal }
\newcommand{\wh}{\widehat}
\newcommand{\wt}{\widetilde}

\newcommand{\eq}[1]{(\ref{#1})}

\begin{document}


\overfullrule=0pt
\parskip=2pt
\parindent=12pt
\headheight=0in \headsep=0in \topmargin=0in \oddsidemargin=0in

\vspace{ -3cm}
\thispagestyle{empty}
\vspace{-1cm}


\begin{center}
\vspace{1cm}
{\Large\bf  
Quantum corrections to short \\ folded superstring in $AdS_{3}\times S^{3}\times M^{4}$
\vspace{1.2cm}
}

\vspace{.2cm}
 {M. Beccaria$^{a}$, 
 G. Macorini$^{b}$
}

\vskip 0.6cm

{\em 
$^{a}$ Dipartimento di Matematica e Fisica ``Ennio De Giorgi'', \\ Universita' del Salento \& INFN, 
                     Via Arnesano, 73100 Lecce, Italy\\
\vskip 0.16cm



$^{b}$  
Niels Bohr International Academy and Discovery Center, Niels Bohr
institute,\\ Blegdamsvej 17 DK-2100 Copenhagen, Denmark


 }

\vspace{.2cm}

\end{center}

\begin{abstract}
We consider integrable superstring theory on $AdS_{3}\times S^{3}\times M^{4}$ where $M^{4}=T^{4}$ or $M^{4}=S^{3}\times S^{1}$ with generic ratio of the radii of the two 3-spheres. We compute the one-loop energy of a short folded string spinning in $AdS_{3}$ and rotating in $S^{3}$.
The computation is performed by world-sheet small spin perturbation theory as well as by quantizing the classical 
algebraic curve characterizing the finite-gap equations. The two methods give equal results up to regularization contributions that are under control. One important byproduct of the calculation is the part of the energy which is due to the dressing phase in the Bethe Ansatz. Remarkably, this contribution $E_{1}^{\rm dressing}$
turns out to be independent on 
the radii ratio. In the $M^{4}=T^{4}$ limit, we discuss how $E_{1}^{\rm dressing}$ relates to a recent proposal
for the dressing phase tested in the $\mathfrak{su}(2)$ sector. We point out some difficulties  suggesting 
that quantization of the $AdS_{3}$ classical finite-gap equations could be subtler than the easier
\ads case.

\end{abstract}

\allowdisplaybreaks

\newpage

\def \os {O(\textstyle{ {1 \ov (\sqrt{\lambda})^2}} )}
\def \ost {O(\textstyle{ {1 \ov (\sqrt{\lambda})^3}} )}
\def \cc {{c }} 
\def \OO {{\cal O}}
\def \te {\textstyle}
\def \fl {\sqrt[4]{\l}}

\def \ha {{{\textstyle{1 \ov2}}}}
\def \fo {{\textstyle{1 \ov4}}}
\def \rx {{\rm x}}
\def \hg {{\hat \g}}

\def \C  {{\rm C}}
\def \hC  {{\rm \hat  C}}
\def \dd  {{\rm d}}
\def \bb {{\rm b}}
\def \dDelta {2}
\def \sql {{\sqrt{\l}}}

 \def \an {{\rm an}} \def \nan {{\rm nan}}
 \def \nm {\tilde n_{11}}
 \def \tn {{\tilde n}}  \def \uni {{\rm inv}}
\def \ttn  {{\bar n}_{11}}
\def \ed {\end{document}}

\def \sj {\te { S\ov J}} 
\def \sjj   {\te { \S\ov \J}}  

\def \ca {{\rm a}} \def \cb {{\rm b}} \def \cc {{a}}  \def \ct {{\rm  v }}  
\def \te {\textstyle} \def \cm {{\rm b}}

\def \ccc {{\tilde a}}

\def \adt {$AdS_3 \times S^3 \times T^4$ }
\def \adst {$AdS_3 \times S^3 \times  S^3 \times S^1$ }

\def \v {h} \def \u {\td h} 
\def \mksl {\mathfrak{sl}(2)}
\def \mksu {\mathfrak{su}(2)}
\def \ccm {{\bar b}}
\tableofcontents

\section{Introduction and summary}

The Maldacena correspondence between quantum strings in \adss  
and $\mathcal N=4$ supersymmetric gauge theory  has been explored in recent years by means of the powerful
unifying framework of integrability \ci{Beisert:2010jr}. 
Integrable structures can be formulated in a non-perturbative way and 
allow to analyze the weak-strong coupling connection in great details. The technical machinery of 
integrability~\footnote{This includes for instance the algebraic curve description of the string classical solutions,
the excitation $S$-matrix, the asymptotic Bethe Ansatz equations and  their TBA extensions for the calculation of 
finite size effects.} is a promising tool to study similar less supersymmetric cases of the duality like
superstring in    $AdS_3 \times   S^3 \times M^4$ 
and $AdS_2 \times   S^2 \times M^6$   supported by R-R fluxes. For these gravitational backgrounds
the dual superconformal theories are poorly understood \cite{deBoer:1999rh,Gukov:2004ym} 
and it is less straightforward to identify the underlying  non-perturbative integrable structures, in particular
the Bethe Ansatz equations.

Recent important progress has been done in the case of strings on $AdS_3 \times   S^3 \times T^4$  and 
   $AdS_3 \times   S^3 \times S^3\times S^1$. They are   described by  the GS superstring 
    action on  the supercosets 
   $PSU(1,1|2) \times PSU(1,1|2)  / SU(1,1) \times SU(2) $   and  
$D(2,1; \a)^{2}  / SU(1,1) \times SU(2)^{2}$.
The  first  model  may be viewed as a special case  of the second.
If the   radius of $AdS_3$ is set to 1, then the radii of the two   3-spheres   can be parametrized as 
$R_1^2 = \a^{-1}, \ R^2_2 = ( 1 - \a)^{-1}$, i.e.  the   $AdS_3 \times   S^3 \times T^4$  model  with $R_2=\infty$ corresponds to $\a=1$.

In \cite{Babichenko:2009dk}, a set of asymptotic Bethe Ansatz (ABA) equations
was proposed for these models starting from the classical    integrable   supercoset  sigma model and 
conjecturing a natural discretisation of the corresponding  finite-gap   equations    following closely  the 
analogy  with  the \adss case \ci{Arutyunov:2004vx}   (see   \cite{Zarembo:2010yz})~\footnote{
We remark that the spectrum described by the finite gap equations is missing two massless modes. One mode corresponds to excitations on $S^{1}$, and the other to a mode shared by the two spheres that is not present in the coset model since the Virasoro constraints are overimposed there. These modes can be put back by hand at the classical level, but it is not yet clear how to do that at the quantum level \cite{Borsato:2012ud,Sax:2012jv}.
In the example discussed in this paper, massless modes cancel out in a conventional world-sheet computation suggesting
that the supercoset description is indeed a consistent truncation like it happens for instance
for the Bethe equations for the $\mathfrak{su}(2)$ sector of AdS5/CFT4 that can be reconstructed from the finite-gap equations on $S^{3}\times \mathbb R^{1}$ \cite{Kazakov:2004qf}, which ignores most part of the string modes in $AdS_{5}\times S^{5}$.
}.
The ABA contains an undetermined dressing phase possibly equal to the BES phase   \cite{Beisert:2006ez}  
appearing in the \adss  case (as well as in $AdS_{4}\times \mathbb{CP}^{3}$ \cite{Gromov:2008qe}).

The ABA system has been analyzed more deeply in in
 \ci{David:2010yg,OhlssonSax:2011ms,Sax:2012jv,Ahn:2012hw}.  In particular, it  was claimed  in  \ci{Ahn:2012hw} that here one cannot
 fix the   dressing  scalar   factors  in the magnon S-matrix  using crossing symmetry  as was  done 
    in the \adss   case \ci{Janik:2006dc,Volin:2009uv,Vieira:2010kb}. This statement does not rule out the simplest
    scenario where the phase is given by the BES expression. However, very recent further developments in \cite{Borsato:2012ud,Borsato:2012ss}
made the story more involved by concluding that there should    be several scalar phase factors   and  that they may differ from the BES  expression. Similar conclusions appeared in \cite{Abbott:2012dd} and point out the necessity of a new phase.

A first study of the specific form of the $AdS_{3}$ phase appeared in \cite{Beccaria:2012kb}
where a proposal (BLMT) has been suggested for the leading 
  quantum string  correction   to the  classical  AFS  phase  in the   ABA  system of  
  \cite{Babichenko:2009dk,OhlssonSax:2011ms}. The BLMT phase was derived by mimicking the analogous steps
  originally applied to the \adss  case  \ci{Beisert:2005cw,Hernandez:2006tk,Freyhult:2006vr,Gromov:2007cd}
  and is based on the study of the phase-dependent ABA predictions to the quantum string  and algebraic curve computations of  the  1-loop  corrections to   semiclassical string   energies~\foot{Previous semiclassical computations for superstrings in $AdS_3 \times S^3 \times M^4$ can be found in \ci{Iwashita:2011ha,Forini:2012bb,Rughoonauth:2012qd,Sundin:2012gc}.}. In particular, the simple example of a rigid circular  string  
  in $AdS_{3}\times S^{3}\subset AdS_{3}\times S^{3}\times T^{4}$
  with two equal spins in $S^3$  \cite{Frolov:2003qc} has been discussed in  \cite{Beccaria:2012kb}, together with the closely related (via  analytic continuation)  case  of $(S,J)$ folded long string \cite{Frolov:2002av}. The conclusion of \cite{Beccaria:2012kb} 
  was a proposal for the phase  in the ABA of \cite{Babichenko:2009dk,OhlssonSax:2011ms}  that is related but differs from the standard  BES form of \ci{Beisert:2005cw,Hernandez:2006tk}.    
  
Unfortunately, the analysis of \cite{Beccaria:2012kb} has some important loose ends. Indeed, the derivation of the phase
requires some {\em ad hoc} steps that work well for the $SU(2)$ circular string, but have an unclear meaning for more
general solutions.
Quantization of any classical string solution amounts to finding the frequencies of the eigenmodes of the classical 
equations of motion, promote them to quantum oscillators with definite frequencies
and sum over the zero point energies $\frac{1}{2}\sum (-1)^{F}\omega$. Frequencies can be found by a standard 
world-sheet analysis or by perturbing the classical algebraic curve. The same set of frequencies should be found, up to 
trivial changes canceling in the sum.
Nevertheless, an infinite summation is involved since each eigenmode has an associated discrete momentum. 
Different prescriptions for the sum can be used. The most natural in the world-sheet approach is the one that knows
nothing about integrability and simply sets a common cut-off on the momentum of each eigenmode. On the other hand, 
the natural prescription when quantizing the algebraic curve is different and assigns a common cut-off on the spectral radius variable associated with each mode. The two prescriptions lead to a finite difference. We shall refer to this
effect naming it a {\em regularization mismatch}. This ambiguity should be fixed, in principle and as usual, 
by fixing finite renormalizations in order to implement the symmetries of the problem. In the problem at hand, symmetries are closely related to the integrable structure and one is led to the hope that properly fixing the latter will
accomplish the desired finite renormalization.

Indeed, this is what happens in the $SU(2)$ circular string \cite{Beccaria:2012kb}. 
Quantization of the algebraic curve
leads to a dressing effect that can be written as an additional one-loop piece $\mc V$ in the finite-gap equations
precisely as in the \ads case \cite{Gromov:2007cd}. However, in $AdS_{3}$, the new piece apparently 
cannot be immediately interpreted as a phase correction in the Bethe equations. Instead $\mc V = \mc V_{\rm phase}+\delta\mc V$, where only $\mc V_{\rm phase}$ admits such an interpretation. Remarkably, 
for the $SU(2)$ circular string, the extra term $\delta\mc V$ happens to cancel exactly the regularization mismatch for yet unclear reasons.
The conclusion is that, for this particular solution, we can enforce dressing effects to be solely encoded in a phase in the 
quantum Bethe equations. This fixes the regularization ambiguity and confirms the world-sheet dressing energy.

In principle, this could be an accident. The discretization/quantization of the unambiguous finite-gap classical string Bethe equations is non trivial. The construction of the discrete all-loop Bethe Ansatz should match the two sides of the AdS/CFT correspondence and there could be 
space for non-trivial new features compared to the $AdS_{5}$ case. Difficulties with the naive quantization are indeed
clearly discussed in \cite{Borsato:2012ss} by comparing the semiclassical finite-gap equations with the near-BMN spectrum. 

The aim of this paper is that of making new steps toward a clarification of this issue by considering a different
classical solution of string theory on $AdS_{3}\times S^{3}\times S^{1}\times S^{1}$ and studying 
its one-loop energy with particular attention to the contributions related to the dressing phase. We focus on 
a folded string spinning in $AdS_{3}$ with semiclassical spin $\S$ and rotating in $S^{3}$ with angular 
momentum $\J$. The short string regime (small $\S$ at generic $\J$) is particularly interesting and much experience is 
known in the analogous $AdS_{5}\times S^{5}$ case \cite{Tirziu:2008fk,Beccaria:2010ry,Gromov:2011bz}
or $AdS_{4}\times \mathbb{CP}^{3}$ \cite{Beccaria:2012qd}. In particular, each term of the small $\S$ expansion of the energy can be considered for large angular momentum $\J$. In this limit, the dressing terms are neatly separated out
and very interesting connections can be studied with the weak-coupling Bethe equations as discussed in 
\cite{Beccaria:2012kp}. 

We study the one-loop energy for a generic $\alpha$-dependent geometry and find that the dressing energy 
is independent on $\alpha$ after the very same redefinition of string tension 
found in \cite{Beccaria:2012kb,Abbott:2012dd} for other classical solution. This redefinition is 
interpreted as a simple rewriting of the energy in terms of the 
interpolating coupling appearing in the Bethe Ansatz equations. The calculation is done
both with world-sheet methods and with quantization of the algebraic curve with perfect agreement. 
Going to the $T^{4}$ limit, where we have the BLMT proposal for the phase, we investigate the interplay of the 
extra term $\delta \mc V$ and the regularization mismatch finding that they do not balance in this case.
This means that we cannot fix the regularization ambiguity in a satisfactory way. The two different $M^{4}=T^{4}$
results described above 
are~\footnote{WS = world-sheet computation with common cut-off on the momenta of all modes, 
AC = algebraic curve quantization with common spectral radius cut-off.}
\ba
\mbox{WS $\equiv$ AC-reg. mismatch}:  && E_{1}^{\rm dressing} = \frac{\coth ^{-1}\left(\sqrt{\mathcal{J}^2+1}\right)}{2 \mathcal{J}^3 \sqrt{\mathcal{J}^2+1}}\,\cS^{2}+\mc O(\S^{3}), \nonumber \\
\mbox{BA with BLMT phase}:  && E_{1}^{\rm dressing} = \bigg[\frac{\coth ^{-1}\left(\sqrt{\mathcal{J}^2+1}\right)}{2 \mathcal{J}^3 \sqrt{\mathcal{J}^2+1}}
+\frac{1}{2\,\J^{4}\,\sqrt{\J^{2}+1}}\bigg]\,\cS^{2}+\mc O(\S^{3}).\nonumber
\ea
Thus, a deeper and more complete
understanding of the role of $\delta \mc V$ is necessary in the general case. This statement can be rephrased by 
saying that at the moment there is no clear matching between the quantization of the finite-gap equations
and the explicit string spectrum. It would be very interesting to test a different set of ABA equations conjectured
for the  $M^{4}=T^{4}$ case like those derived in \cite{Borsato:2012ud,Borsato:2012ss} for $0<\alpha<1$.

\bigskip

The outline of the paper is the following. In Sec.~(\ref{sec:folded}) we derive the one-loop energy 
for the folded string in $\alpha$-dependent $AdS_{3}\times S^{3}\times S^{1}\times S^{1}$ by world-sheet
perturbation theory. In Sec.~(\ref{sec:AC}), we derive the same set of frequencies by quantizing the classical
algebraic curve and discuss the relation between the two approaches by computing the regularization mismatch.
In Sec.~(\ref{sec:Bethe}), we match the large $\J$ expansion of the string dressing energy from the weak-coupling 
large $J = \sqrt\lambda\,\J$ expansion of the one and two-loop.
Finally, in Sec.~(\ref{sec:Dressing}), we explore the possibility of recovering the dressing one-loop energy from 
a suitable dressing phase in the Bethe equations.
Various appendices are devoted to technical details.

\section{Folded string in $AdS_{3}\times S^{3}\times S^{1}\times S^{1}$}
\label{sec:folded}

\subsection{Classical solution}

The classical solution we are going to consider is described in Appendix C
of \cite{Forini:2012bb} and has the same form as the analogous folded string solution in 
$AdS_{5}\times S^{5}$ \cite{Frolov:2002av}.
The metric of $AdS_{3}\times S^{3}\times S^{3}\times S^{1}$ is parametrized as
\ba
ds^{2} &=& R^{2}\,\bigg(ds^{2}_{\rm AdS}+\frac{1}{\alpha}ds^{2}_{S^{3}_{+}}+\frac{1}{1-\alpha}ds^{2}_{S^{3}_{-}}\bigg)+dU^{2}, \\
ds^{2}_{\rm AdS} &=& -\cosh^{2}\rho\,dt^{2}+d\rho^{2}+\sinh^{2}\rho\,d\phi^{2}, \\
ds^{2}_{S^{3}_\pm} &=& d\beta_{\pm}^{2}+\cos^{2}\beta_{\pm}(d\gamma_{\pm}^{2}+\cos^{2}\gamma_{\pm}d\varphi_{\pm}^{2}).
\ea
The relation between the Anti de Sitter radius $R$ and the radii of the two $S^{3}$ spheres is a consequence of the supergravity equations of motion.

The classical folded string solution follows from  the Ansatz
\ba
&& t = \kappa\,\tau, \quad \phi = w\,\tau, \quad \rho(\sigma)=\rho(\sigma+2\pi), \\
&& \varphi_{\pm} = \J_{\pm}\,\tau, \quad \gamma_{\pm}=\beta_{\pm}=U=0.
\ea
We shall assume the following distribution of the sphere angular momenta $\J_{\pm}$ between the two spheres
\be
\J_{+} = \alpha\,\J, \qquad \J_{-} = (1-\alpha)\,\J.
\ee
This choice amounts to have a well-defined BPS limit for $\S=0$.
The equation of motion for $\rho(\sigma)$ and the Virasoro constraint takes the same form as in $AdS_{5}\times S^{5}$. In particular, we have 
\be
\rho'^2 = \kappa^2 \cosh^2 \rho - w^2 \sinh^2 \rho -\J^2 .   
\ee
The coordinate 
$\rho$ varies from $0$ to its maximal  value $\rho_*$
\begin{equation}
\coth^2 \rho_* = \frac{w^2-\J^2}{\kappa^2-\J^2}\equiv 1+ \frac{1}{\varepsilon^2},
\ee
where $\varepsilon$ measures the length of the string and is small in the short string limit.
The solution of the  differential equation for $\rho$, {\em i.e.}
\begin{equation}
\rho' = \pm \sqrt{\kappa^2-\J^2} \sqrt{1-\eps^{-2}\, {\sinh^2 \rho}}\ , \quad \quad \rho(0)=0,
\end{equation}
 can be written in terms of a  Jacobi elliptic function
\begin{equation}
\sinh \rho=\eps\, \ {\rm sn}({\sqrt{\kappa^2-\J^2}\,\eps^{-1}\,  \sigma},\ -\eps^2) \ .
\ee
The periodicity condition and the charges are
 \ba
&& \sqrt{\kappa^2-\J^2}=\eps \
 _2 F_1\left(\frac{1}{2},\frac{1}{2};1;-\eps^2\right) \ , \  \ \ \  \
  \ \ \mathcal{E}_0\equiv \frac{E_0}{\sqrt{\lambda}}=\frac{\kappa}{\sqrt{\kappa^2-\J^2}}
\eps \
  _2F_1\left(-\frac{1}{2},\frac{1}{2};1;-\eps^2\right) , \nonumber \\
&&
 \mathcal{S}\equiv \frac{S}{\sqrt{\lambda}}=\frac{w}{\sqrt{\kappa^2-\J^2}}\frac{\eps^3}{2}
 \ _2F_1\left(\frac{1}{2},\frac{3}{2};2;-\eps^2\right) \ , \quad \quad J= \J \sqrt{\lambda} \ .
\ea

\subsection{One-loop correction to the energy: World-sheet computation}
\label{sec:method}

The one-loop energy is obtained according to the following general recipe. First, we compute the Lagrangian
for the quadratic fluctuations by shifting all fields $\Phi = (t, \phi, \rho, \dots)$ with respect to the their classical 
values and evaluating the action at quadratic level
\be
\mathscr S[\Phi] = \mathscr S\bigg[\Phi_{classical}+\frac{1}{\sqrt\lambda}\widetilde \Phi\bigg] = \mathscr S[\Phi_{classical}]
+\int d\tau d\sigma\,\widetilde \Phi^{T}\,\mathscr D(\partial_{\sigma}, \partial_{\tau}, \sigma)\,\wt \Phi+\dots.
\ee
Then, for each integer mode number $n$, we consider a universal time dependence $e^{i\,\omega_{n}\,\tau}$
and find the perturbative solution to the equation 
\be
\mathscr D(\partial_{\sigma}, i\,\omega_{n}, \sigma)\,\widetilde \Phi_{n}^{I}(\sigma) = 0,
\ee
where
\ba
\widetilde\Phi_{n}^{I} &=& e^{i\,n\,\sigma}\,\widetilde \Phi_{n}^{I\,(0)}+\eps\,\widetilde\Phi^{I\,(1)}_{n}(\sigma)+\eps^{2}\,\widetilde\Phi^{I\,(2)}_{n}(\sigma)+\dots, \\
\omega_{n}^{I} &=& \omega^{I\,(0)}_{n}+\eps\, \omega^{I\,(1)}_{n}+\eps^{2}\, \omega^{I\,(2)}_{n}+\dots.
\ea 
This expansion has a solution for certain constant vectors $\wt \Phi_{n}^{I\,(0)}$ that are associated, as $I$ varies, to 
the various fields of the problem. The other coefficient functions $\wt \Phi_{n}^{I\,(1, 2, \dots)}$
follow from the equation. Finally, we obtain the one-loop energy, but summing over the properly normalized 
zero point energies 
\be
E_{1} = \frac{1}{2\,\kappa}\sum_{I}(-1)^{F_{I}}\sum_{n\in\mathbb Z}\omega_n^{I} \equiv \frac{1}{2\,\kappa}\,\mathfrak S,
\ee
where $(-1)^{F_{I}}$ is the Bose-Fermi sign of the field. For later use, we have denoted by $\mathfrak S$ the signed sum over frequencies without the factor $1/(2\,\kappa)$.

\subsubsection{Bosonic quadratic fluctuations}

The bosonic fluctuations are discussed in \cite{Forini:2012bb} in the static gauge. Denoting by a tilde the fluctuations, 
the static gauge fixes $\wt t = \wt \rho=0$. Introducing the linear combinations 
\be
\varphi = \varphi_{+}+\varphi_{-}, \qquad \psi = -\sqrt\frac{1-\alpha}{\alpha}\,\varphi_{+}+\sqrt\frac{\alpha}{1-\alpha}\,\varphi_{-},
\ee
there are then two coupled fields $(\wt \phi, \wt\varphi)$, 
two massless fields $\wt \psi$, $\wt U$, and the decoupled massive fields $\wt\beta_{\pm}$, $\wt\gamma_{\pm}$
with equal masses $\J_{\pm}$.
The coupled sector in the static gauge is described by~\footnote{Notice that for quadratic fluctuations the Pohlmeyer reduction \cite{Iwashita:2010tg} (absorbing a global factor 2 in the normalization of the fields) gives the same result.}
\ba
\mathcal L_{\rm coupled} &=& -\partial^{a}a_{3}\partial_{a}a_{3}-m_{3}^{2}\,a_{3}^{2} -\partial^{a}a_{4}\partial_{a}a_{4}-m_{4}^{2}\,a_{4}^{2}+\frac{4w\kappa\J}{\J^{2}+\rho'^{2}}\,a_{4}\,\partial_{\tau}a_{3},
\ea
where 
\ba
m_{3}^{2} &=& \J^{2}+2\rho'^{2}
+\frac{2w^{2}\kappa^{2}}{\J^{2}+\rho'^{2}}-\frac{3w^{2}\kappa^{2}\J^{2}}{(\J^{2}+\rho'^{2})}, \\
m_{4}^{2} &=& \J^{2}\bigg[-1+\frac{2(w^{2}+\kappa^{2})}{\J^{2}+\rho'^{2}}-\frac{3w^{2}\kappa^{2}}{(\J^{2}+\rho'^{2})^{2}}\bigg].
\ea
Thus, $\mathcal L_{\rm coupled} = (a_{3}, a_{4})\,Q_{B}\,(a_{3}, a_{4})^{T}$, with 
\be
Q_{B} = \left(\begin{array}{cc}
 \partial^{a}\partial_{a}-m_{3}^{2} & \displaystyle -\frac{2w\kappa\J}{\J^{2}+\rho'^{2}}\,\partial_{\tau}\\ \\
\displaystyle \frac{2w\kappa\J}{\J^{2}+\rho'^{2}}\,\partial_{\tau}& \partial^{a}\partial_{a}-m_{4}^{2} 
\end{array}
\right).
\ee

\subsubsection{Fermionic quadratic fluctuations}

The  operator that describes the quadratic fermionic fluctuations is 
\ba
\label{eq:fermionic}
 \mathcal D_{F} &=& \Gamma^{a}\,\partial_{a}-\frac{\kappa\,w\,\J}{2\,(\rho'^{2}+\J^{2})}\Gamma^{12}\overline\Gamma+\frac{\rho'}{2}(\Gamma^{012}-\sqrt\alpha\,\Gamma^{345}-\sqrt{1-\alpha}\,\Gamma^{678})\nonumber\\
&& +\frac{\sqrt{\rho'^{2}+\J^{2}}-\rho'}{2}\,(\Gamma^{012}-(\alpha\,\Gamma^{34}+(1-\alpha)\,\Gamma^{67})\,\overline\Gamma),
 \ea
where 
\be
 \overline\Gamma = \sqrt\alpha\,\Gamma^{5}+\sqrt{1-\alpha}\,\Gamma^{8}.
\ee
This operator is rather complicated, but can be simplified as explained in App.~(\ref{app:fermionic}).
 
\subsection{Short string expansion of frequencies in the flat space limit}

\noindent{\bf Bosons}
\medskip
 
It is convenient to rotate the $Q_{B}$ operator as 
\be
Q_{B} \to R^{-1}_{B}\,Q_{B}\,R_{B},\qquad R_{B} = \left(
\begin{array}{cc}
  \frac{1}{2} & \frac{1}{2} \\
  -\frac{i}{2} & \frac{i}{2} \\
\end{array}
\right).
\ee
Then $Q_{B}=Q_{B}^{(0)}+\eps Q_{B}^{(1)}+\dots$ and 
\be
Q_{B}^{(0)} = \left(
\begin{array}{cc}
 -n^2+\omega ^2-2 \sqrt{\mathcal{J}^2+1} \omega -1 & 0 \\
  0 & -n^2+\omega ^2+2 \sqrt{\mathcal{J}^2+1} \omega -1 \\
\end{array}
\right),
\ee
when acting on functions $\sim e^{in\sigma}$. At finite $\J>0$, the eigenvalues of $Q_{B}^{(0)}$ can be written as  
\be
\pm(\sqrt{n^{2}+\J^{2}}+\sqrt{1+\J^{2}}),\qquad \pm (\sqrt{n^{2}+\J^{2}}-\sqrt{1+\J^{2}}) .
\ee
Hence, the coupled system contributes the following set of $\omega>0$ frequencies
\be
\sqrt{n^{2}+\J^{2}}\pm\sqrt{1+\J^{2}}.
\ee
To these two frequencies, we have to add the two massless and two massive decoupled contributions that read
\be
2\times\left\{n, \quad \sqrt{n^{2}+\alpha^{2}\J^{2}}, \quad \sqrt{n^{2}+(1-\alpha)^{2}\J^{2}}
\right\}.
\ee

{\bf Fermions}
\medskip

Evaluating at $\eps=0$ the frequencies coming from the 8 blocks in which we decomposed $-\mathcal D_{F}^{2}$
(see App.~(\ref{app:fermionic})),
we find after some calculation the following fermionic frequencies (plus the opposite one due to the symmetry 
$\omega\to -\omega$)
\ba
&& n\pm\frac{1}{2}\sqrt{\J^{2}+1}, \\
&& \sqrt{n^{2}+\J^{2}}\pm\frac{1}{2}\sqrt{\J^{2}+1}, \\
&& \sqrt{n^{2}+\alpha^{2}\J^{2}}\pm\frac{1}{2}\sqrt{\J^{2}+1}, \\
&& \sqrt{n^{2}+(1-\alpha)^{2}\J^{2}}\pm\frac{1}{2}\sqrt{\J^{2}+1} .
\ea

{\bf Summary}
\medskip

In summary, in the flat space limit $\eps=0$, we have the following contributions from the various bosonic fields

\be
\begin{array}{ccc}
\rm multiplicity & \rm field(s) & \omega_{n} \\ 
\hline\hline\\
1 & (\phi, \varphi) & \sqrt{n^{2}+\J^{2}}+\sqrt{\J^{2}+1} \\
&& \sqrt{n^{2}+\J^{2}}-\sqrt{\J^{2}+1} \\ \\
2 & U, \psi & n \\ \\
2 & \beta_{\pm}, \gamma_{\pm} & \sqrt{n^{2}+\alpha^{2}\J^{2}} \\ 
&& \sqrt{n^{2}+(1-\alpha)^{2}\J^{2}} \\ \\ \hline\hline
\end{array}
\ee
as well as fermionic ones
\be
\begin{array}{ccc}
\rm multiplicity & \rm field(s) & \omega_{n} \\ 
\hline\hline\\
1 & \Psi &  \sqrt{n^{2}+\J^{2}}+\frac{1}{2}\,\sqrt{\J^{2}+1} \\
&& \sqrt{n^{2}+\J^{2}}-\frac{1}{2}\,\sqrt{\J^{2}+1}  \\
&& n+\frac{1}{2}\,\sqrt{\J^{2}+1}  \\
&& n-\frac{1}{2}\,\sqrt{\J^{2}+1}  \\
&& \sqrt{n^{2}+\alpha^{2}\J^{2}}+\frac{1}{2}\,\sqrt{\J^{2}+1}  \\
&& \sqrt{n^{2}+(1-\alpha)^{2}\J^{2}}-\frac{1}{2}\,\sqrt{\J^{2}+1}  \\ \\ \hline\hline
\end{array}
\ee
\bigskip
Summing with weight $(-1)^{F}$ we find full cancellation of all terms. Notice that 
there are no ghost since we are in the static gauge.

\subsection{$\mc O(\S)$ frequencies and one-loop energy}

Computing the $\mc O(\eps^{2}\sim \S)$
corrections to the various frequencies we find the following results

\medskip
{\bf Bosons}
\medskip

For the bosonic modes, we have two massless modes and 2+2 massive decoupled modes with masses
$\alpha\,\J$, $(1-\alpha)\,\J$ that do not receive corrections
\ba
\omega_{n}^{B\,(1,2)} &=& n, \\
\omega_{n}^{B\,(3,4)} &=& \sqrt{n^{2}+\alpha^{2}\J^{2}}, \\
\omega_{n}^{B\,(5,6)} &=& \sqrt{n^{2}+(1-\alpha)^{2}\J^{2}}, 
\ea
The two coupled modes give instead
\be
\omega_{n}^{B\,(7,8)} = \sqrt{n^{2}+\J^{2}}\pm\sqrt{\J^{2}+1}+\eps^{2}\,\bigg(\frac{1}{2\sqrt{\J^{2}+n^{2}}}\pm\frac{1}{4\sqrt{\J^{2}+1}}\bigg)+\dots .
\ee

\medskip
{\bf Fermions}
\medskip

The corrections to the fermionic frequencies are
{\small
\ba
\omega_{n}^{F\,(1,2)} &=& n\pm\frac{\sqrt{\J^{2}+1}}{2}\pm\eps^{2}\,\frac{1}{8\,\sqrt{\J^{2}+1}}+\dots,
 \\ \nonumber \\
\omega_{n}^{F\,(3,4)} &=& \sqrt{n^{2}+\J^{2}}\pm\frac{\sqrt{\J^{2}+1}}{2}
+\eps^{2}\,\bigg(
\frac{1}{4\,\sqrt{\J^{2}+n^{2}}}\pm\frac{1}{8\,\sqrt{\J^{2}+1}}
\bigg) +\dots,\\  \nonumber \\
\omega_{n}^{F\,(5,6)} &=& \sqrt{n^{2}+\alpha^{2}\J^{2}}\pm\frac{\sqrt{\J^{2}+1}}{2}+\eps^{2}\,\bigg(
\frac{\alpha}{4\,\sqrt{\alpha^{2}\J^{2}+n^{2}}}\pm\frac{1}{8\,\sqrt{\J^{2}+1}}
\bigg)+\dots,\\  \nonumber \\
\omega_{n}^{F\,(7,8)} &=& \sqrt{n^{2}+(1-\alpha)^{2}\J^{2}}\pm\frac{\sqrt{\J^{2}+1}}{2}+\eps^{2}\,\bigg(
\frac{1-\alpha}{4\,\sqrt{(1-\alpha)^{2}\J^{2}+n^{2}}}\pm\frac{1}{8\,\sqrt{\J^{2}+1}}
\bigg) +\dots .
\ea
}
Notice that they have the general form 
\be
\omega^{F}_{n} = \sqrt{n^{2}+\xi^{2}\J^{2}}\pm\frac{\sqrt{\J^{2}+1}}{2}+\eps^{2}\,\bigg(
\frac{\xi}{4\,\sqrt{\xi^{2}\J^{2}+n^{2}}}\pm\frac{1}{8\,\sqrt{\J^{2}+1}}
\bigg)+\dots,
\ee
where $\xi = 0, 1, \alpha, 1-\alpha$.
If we combine bosons and fermions for generic $n$ we find 
\be
\sum_{i=1}^{8}(\omega_{n}^{B\,(i)}-\omega_{n}^{F\,(i)}) = \frac{\eps^{2}}{2}\bigg(
\frac{1}{\sqrt{n^{2}+\J^{2}}}-\frac{\alpha}{\sqrt{n^{2}+\alpha^{2}\J^{2}}}
-\frac{1-\alpha}{\sqrt{n^{2}+(1-\alpha)^{2}\J^{2}}}
\bigg)+\dots, 
\ee

\medskip
{\bf Low special modes}
\medskip

We have to take special care of low modes due to possible resonances with the $\sigma$ dependent terms in the various differential operators that govern the quadratic fluctuations. This is signaled by divergent terms in the eigenfunctions
or eigenvalues at special values of $n$. 
The fermionic frequencies do not have special low modes with the exception of 
\be
\omega_{n}^{F\,(4)} = \sqrt{n^{2}+\J^{2}}-\frac{\sqrt{\J^{2}+1}}{2}
+\eps^{2}\,\bigg(
\frac{1}{4\,\sqrt{\J^{2}+n^{2}}}-\frac{1}{8\,\sqrt{\J^{2}+1}}
\bigg) +\dots,
\ee
and its analogues with $\alpha$ dependence. For $n=1$ there are singularities in the wave function. Doing a more careful computation for the values $n=\pm 1$ we find that these modes mix and lead to two correction whose sum is twice the naive above value. So, in the calculation of the energy that is summed over $n$, we can use the above expressions.
Bosons are special only in the case $\omega^{B\,(8)}_{\pm 1}$. These two frequencies are exactly zero at order $\eps^{2}$ included.

\medskip
{\bf One-loop energy at order $\mc O(\S)$}
\medskip

In summary, the relevant sum of frequencies is 
\ba
\mathfrak S = \eps^{2}\,\mathfrak S^{(2)} = \sum_{i=1}^{8}(\omega_{0}^{B\,(i)}-\omega_{0}^{F\,(i)}) +2\,\bigg[-\omega_{1}^{B\,(8)}+
\sum_{n=1}^{\infty}\sum_{i=1}^{8}(\omega_{n}^{B\,(i)}-\omega_{n}^{F\,(i)}) 
\bigg],
\ea
where the generic $n$ expressions for the $\omega$'s have to be used and the subtraction in square brackets takes into 
account that the true value of $\omega^{B\,(8)}_{\pm 1}$ is zero, as discussed above. We can add and subtract in order to write
\be
\mathfrak S^{(2)} = \alpha\,\log\alpha+(1-\alpha)\,\log(1-\alpha)-\frac{1}{2\,\sqrt{\J^{2}+1}}+\Delta \mathfrak S^{(2)},
\ee
where
\ba
\Delta \mathfrak S^{(2)} &=& 2\,
\sum_{n=1}^{\infty}\sum_{i=1}^{8}(\omega_{n}^{B\,(i)}-\omega_{n}^{F\,(i)}) - \alpha\,\log\alpha-(1-\alpha)\,\log(1-\alpha)-\frac{1}{2\,\J} = \\
&=& - \alpha\,\log\alpha-(1-\alpha)\,\log(1-\alpha)-\frac{1}{2\,\J}\nonumber \\
&& +\sum_{n=1}^{\infty}\bigg[\frac{1}{\sqrt{n^{2}+\J^{2}}}-\frac{\alpha}{\sqrt{n^{2}+\alpha^{2}\J^{2}}}
-\frac{1-\alpha}{\sqrt{n^{2}+(1-\alpha)^{2}\J^{2}}}\bigg].\nonumber
\ea
Remarkably, the quantity $\Delta \mathfrak S^{(2)}$ {\bf is exponentially small } for large $\J$. The one-loop energy is obtained by dividing by $2\kappa$. 
Taking into account that $\kappa = \J + \dots$ and that $\eps^{2} = \frac{2\cS}{\sqrt{\J^{2}+1}}+\dots$, we thus find 
\be
E_{1} = \cS\,\bigg[
\frac{\alpha\,\log\alpha+(1-\alpha)\,\log(1-\alpha)}{\J\,\sqrt{\J^{2}+1}}-\frac{1}{2\,(\J+\J^{3})}
\bigg]+{\rm exponentially\ suppressed\ at\ large\ \J}
\ee
The second term in square bracket is the same as in $AdS_{5}\times S^{5}$ \cite{Gromov:2011bz}
and is the one-loop term of the exact slope function derived in \cite{Basso:2011rs, Gromov:2012eg} (see also 
\cite{Gromov:2011bz}).
The first term can be removed by a coupling redefinition as explained in \cite{Beccaria:2012kb}.
If we shift the string tension as
\be
\label{shiftf}
	\sqrt\lambda\to \sqrt\lambda-4\,\pi\,a,
\ee
while holding the charges $S$ and $J$ fixed, then the one-loop energy will get a contribution coming from the classical energy $E_{0}$. Then choosing
\be
	a=\frac{\alpha\,\log\alpha+(1-\alpha)\,\log(1-\alpha)}{4\pi},
\ee
we see that the $\alpha$-dependent term in terms with $L$ is removed. The shift \eq{shiftf} is equivalent to rewriting the string result in terms of the  interpolating coupling discussed in  \cite{Beccaria:2012kb}
for the $SU(2)$ circular string as well as for the long folded string
\be
\label{shiftfbis}
	h(\lambda)=\frac{\sqrt\lambda}{4\pi}+{ \alpha\,\log\alpha+(1-\alpha)\,\log(1-\alpha)\ov 4 \pi} +\O\big({ 1 \ov \sql} \big).
\ee	
In Sec.(\ref{sec:Bethe}), we will see that it is natural to identify $h(\lambda)$ with the Bethe Ansatz interpolating coupling.

\bigskip
Altough we are mainly interested in string theory on $AdS_{3}\times S^{3}\times M^{4}$, we emphasize that 
the frequency method discussed in this section can of course be applied to the study of the short folded string 
in  $AdS_{5}\times S^{5}$. This application does not lead to any new results, but allows a reconciliation
of the results presented in \cite{Tirziu:2008fk,Beccaria:2010ry}
with the exact slope prediction \cite{Basso:2011rs, Gromov:2012eg,Gromov:2011bz}. The agreement was not possible in previous papers
due to the choice $\J=0$ and to the fact that the fermionic quadratic fluctuation operator has to be modified
as discussed in \cite{Iwashita:2010tg,Forini:2012bb}. Details of the $AdS_{5}\times S^{5}$ application are collected
in App.~(\ref{app:ads5-application}).

\subsection{$\mc O(\S^{2})$ frequencies and one-loop energy}

After a long computation, we find the following non-trivial corrections

\medskip
{\bf Bosons}
\medskip

The two coupled modes give 
\ba
\omega_{n}^{B\,(7,8)} &=& \sqrt{n^{2}+\J^{2}}\pm\sqrt{\J^{2}+1}+\eps^{2}\,\bigg(\frac{1}{2\sqrt{\J^{2}+n^{2}}}\pm\frac{1}{4\sqrt{\J^{2}+1}}\bigg)\nonumber \\
&&+\eps^{4}\,\bigg(\pm
\frac{-5 \mathcal{J}^4-7 \mathcal{J}^2-\left(9 \mathcal{J}^4+15 \mathcal{J}^2+4\right) n^4+2 \left(11 \mathcal{J}^4+19 \mathcal{J}^2+6\right) n^2}{64 \mathcal{J}^2 \left(\mathcal{J}^2+1\right)^{3/2}
   \left(n^2-1\right)^2}\nonumber \\
   &&+\frac{-2 n^4+\mathcal{J}^4 \left(-3 n^4+5 n^2-4\right)-\mathcal{J}^2 \left(3 n^6-3 n^4+2 n^2+2\right)}{16 \mathcal{J}^2 \left(n^2-1\right)^2 \left(\mathcal{J}^2+n^2\right)^{3/2}}
\bigg)+\dots.
\ea

\medskip
{\bf Fermions}
\medskip

The $\mc O(\eps^{4})$ corrected fermionic frequencies can be compactly written as 
\ba
\omega_{n}^{F\,(1,2)} &=& \omega_{n}^{F}(0,\pm 1), \\
\omega_{n}^{F\,(3,4)} &=& \omega_{n}^{F}(1,\pm 1), \\
\omega_{n}^{F\,(5,6)} &=& \omega_{n}^{F}(\alpha,\pm 1), \\
\omega_{n}^{F\,(7,8)} &=& \omega_{n}^{F}(1-\alpha,\pm 1), 
\ea
where
{\small
\ba
 && \omega_{n}^{F}(\xi,\sigma) = \sqrt{n^{2}+\xi^{2}\,\J^{2}}+\frac{\sigma}{2}\,\sqrt{\J^{2}+1}+
\eps^{2}\,\bigg(
\frac{\xi}{4\,\sqrt{\xi^{2}\J^{2}+n^{2}}}+\sigma\frac{1}{8\,\sqrt{\J^{2}+1}}
\bigg)\\
&& \!\!\!\!\!\!\!+\eps^{4}\bigg[
-\frac{\xi  \left(2 \mathcal{J}^2 \left(2 \mathcal{J}^2+1\right) \xi ^6+3 \mathcal{J}^2 n^6+\xi  n^4 \left(3 \mathcal{J}^4 \xi
   +\mathcal{J}^2 (3-9 \xi )-4 \xi +3\right)+\xi ^3 n^2 \left(\mathcal{J}^4 (2-9 \xi )+\mathcal{J}^2+2 \xi -1\right)\right)}{32
   \mathcal{J}^2 \left(n^2-\xi ^2\right)^2 \left(\mathcal{J}^2 \xi ^2+n^2\right)^{3/2}} \nonumber \\
   &&
 \!\!\!\!\!\!\!+\sigma\,   \frac{ \left(-\mathcal{J}^2 \left(5 \mathcal{J}^2+7\right) \xi ^4-\left(9 \mathcal{J}^4+15 \mathcal{J}^2+4\right) n^4+2
   \xi  n^2 \left(\mathcal{J}^4 (3 \xi +4)+\mathcal{J}^2 (3 \xi +8)-2 \xi +4\right)\right)}{128 \mathcal{J}^2
   \left(\mathcal{J}^2+1\right)^{3/2} \left(n^2-\xi ^2\right)^2}\bigg]+\dots.\nonumber
\ea
}

\medskip
{\bf Low special modes}
\medskip

The low modes can be discussed as in the previous section. In particular, we find that at this order it is still true that
$\omega_{\pm 1}^{B\,(8)}=0$. There is also one fermionic frequency that deserves a special analysis. It is 
$\omega_{n}^{F\,(4)}$ for $n=\pm 1$.  The analysis of this special case shows that 
\be
\omega_{1}^{F\,(4)}+\omega_{-1}^{F\,(4)} =\sqrt{\mathcal{J}^2+1}+\frac{\epsilon ^2}{4 \sqrt{\mathcal{J}^2+1}}+\frac{\left(-5 \mathcal{J}^2-7\right) \epsilon
   ^4}{64 \left(\mathcal{J}^2+1\right)^{3/2}}+\dots.
\ee
All other cases are not special in any sense and the above generic-$n$ expressions can be used.

\medskip
{\bf One-loop energy at order $\mc O(\S^{2})$}
\medskip

The sum over frequencies is now 
\be
\mathfrak S = \eps^{2}\,\mathfrak S^{(2)}+\eps^{4}\,\mathfrak S^{(4)}+\dots. 
\ee
The term $\mathfrak S^{(4)}$ can be further split into a part which is exponentially suppressed for large $\J$ and a remainder
\be
\mathfrak S^{(4)} = \mathfrak S^{(4)\, \rm wrap}+\mathfrak S^{(4)\, \rm non-wrap}.
\ee 
The splitting is achieved by the methods illustrated in App.~(\ref{app:sums}). 
The one-loop energy is obtained from $E_{1} = \frac{S}{2\kappa}$, with
\ba
\kappa &=& \mathcal{J}+\frac{\mathcal{S}}{\mathcal{J} \sqrt{\mathcal{J}^2+1}}-\frac{\left(3 \mathcal{J}^4+7 \mathcal{J}^2+2\right)
   \mathcal{S}^2}{4 \left(\mathcal{J}^3 \left(\mathcal{J}^2+1\right)^2\right)}\nonumber \\
   && +\frac{\left(12 \mathcal{J}^8+51 \mathcal{J}^6+77
   \mathcal{J}^4+36 \mathcal{J}^2+8\right) \mathcal{S}^3}{16 \mathcal{J}^5
   \left(\mathcal{J}^2+1\right)^{7/2}}+O\left(\mathcal{S}^4\right), \\
   \eps^{2} &=& \frac{2 \mathcal{S}}{\sqrt{\mathcal{J}^2+1}}+\frac{\left(\mathcal{J}^2-1\right) \mathcal{S}^2}{2
   \left(\mathcal{J}^2+1\right)^2}+\frac{\left(-2 \mathcal{J}^4-5 \mathcal{J}^2+7\right) \mathcal{S}^3}{8
   \left(\mathcal{J}^2+1\right)^{7/2}}+O\left(\mathcal{S}^4\right).
\ea 
The non-wrapping part of $E_{1}$ turns out to be 
\ba
\lefteqn{E_{1}^{\rm non-wrap} = \bigg[
\frac{L}{\J\,\sqrt{\J^{2}+1}}-\frac{1}{2\,(\J+\J^{3})}
\bigg]\,\cS} && \\
&& +\bigg[
\bigg(\frac{\mathcal{J}}{2 \left(\mathcal{J}^2+1\right)^2}-\frac{1}{\mathcal{J}^3}\bigg)\,L +
\frac{1}{48 \mathcal{J}^3 \left(\mathcal{J}^2+1\right)^{5/2}}\bigg(
6 \mathcal{J}^4+12 \pi  \left(\mathcal{J}^2+1\right)^2 \left((2 \alpha -1) \cot (\pi  \alpha ) \right. \nonumber \\
&& \left. -\pi  (\alpha -1) \alpha  \csc
   ^2(\pi  \alpha )\right)+45 \mathcal{J}^2-4 \pi ^2 \left(\mathcal{J}^2+1\right)^2+12 \left(\mathcal{J}^2+1\right)^2 \coth
   ^{-1}\left(\sqrt{\mathcal{J}^2+1}\right)+12\bigg)
\bigg]\,\cS^{2}+\dots ,\nonumber
\ea
where $L=\alpha\log(\alpha)+(1-\alpha)\log(1-\alpha)$.

\subsection{Large $\J$ expansion and non-analytic contributions}

Expanding at large $\J$, where $E_{1} = E_{1}^{\rm non wrap}$ up to exponentially small corrections,  we find
\ba
E_{1} &=& \bigg[
\frac{L}{\mathcal{J}^2}-\frac{1}{2 \mathcal{J}^3}-\frac{L}{2 \mathcal{J}^4}+\frac{1}{2 \mathcal{J}^5}+\frac{3 L}{8
   \mathcal{J}^6}-\frac{1}{2 \mathcal{J}^7}+\dots
\bigg]\,\cS\\
&&+\bigg[-\frac{L}{2\,\J^{3}}+\frac{F(\alpha)}{\J^{4}}+\frac{\frac{1}{4}-L}{\J^{5}}
+\frac{-\frac{1}{2}\,F(\alpha)+\frac{11}{16}}{\J^{5}}+\frac{\frac{3}{2}\,L-\frac{1}{6}}{\J^{7}}+\dots
\bigg]\,\cS^{2}+\dots, \nonumber
\ea
where  the function $F(\alpha)$ is 
\be
\label{eq:alpha}
F(\alpha) = \frac{1}{4} \pi ^2 (1-\alpha ) \alpha  \left(\cot ^2(\pi  \alpha
   )+1\right)+\frac{1}{4} \pi  (2 \alpha -1) \cot (\pi  \alpha )-\frac{\pi
   ^2}{12}+\frac{1}{8}.
\ee
Remarkably, all the $L$-dependent terms can be removed by the previous coupling redefinition (\ref{shiftfbis}).
In the following, we shall systematically drop them since our aim will be that of comparing the world-sheet 
computation with other methods based on integrability (Algebraic Curve quantization, discrete Bethe equations).
In summary, the large $\J$ expansion 
of the classical and one-loop energies can be written as 
\ba
\mc E_{0} &=&  = \mathcal{J}+
\frac{\sqrt{\mathcal{J}^2+1}}{\mathcal{J}}\, \mathcal{S}+
\frac{\left(-\mathcal{J}^2-2\right)}{4
   \mathcal{J}^3 \left(\mathcal{J}^2+1\right)}\, \mathcal{S}^2+\dots \\
   &=& \mc J + \bigg(
   1
   +\frac{1}{2 \mathcal{J}^2}
   {-\frac{1}{8 \mathcal{J}^4}}
   {+\frac{1}{16 \mathcal{J}^6}}
   +\dots
   \bigg)\,\mc S+\bigg(
   -\frac{1}{4 \mathcal{J}^3}
   {-\frac{1}{4 \mathcal{J}^5}}
   {+\frac{1}{4 \mathcal{J}^7}}
   +\dots
   \bigg)\,\mc S^{2}+\dots, \nonumber \\
E_{1} &=& \bigg(
-\frac{1}{2\J^{3}}
{+\frac{1}{2\J^{5}}}
{-\frac{1}{2\J^{7}}}
+\dots
\bigg)\,\cS+\bigg(
\frac{F(\alpha)}{\J^{4}}
{+\frac{1}{4\,\J^{5}}}
{+\frac{-\frac{1}{2}\,F(\alpha)+\frac{11}{16}}{\J^{6}}}
{-\frac{1}{6\J^{7}}}
+\dots
\bigg)\,\cS^{2}+\dots\ \ . \nonumber
\ea
It is convenient to regroup the various terms appearing in the large $\J$ expansion in a way that will ease
the comparison with the weak-coupling expansion of the Bethe 
equations after re-expansion in powers of $\lambda$ at fixed $J$ as we shall discuss in in Sec.~(\ref{sec:Bethe}).
Thus, we rewrite the above expansions as 
\ba
\label{eq:string0}
\mc E_{0} &=& \bigg[\mc J + \bigg(
   1
   +\frac{1}{2 \mathcal{J}^2}\bigg)\,\S
   -\frac{1}{4 \mathcal{J}^3}\,\S^{2}+\dots\bigg]_{1L}
   +\bigg[-\frac{1}{8 \mathcal{J}^4}\,\S
   -\frac{1}{4 \mathcal{J}^5}\,\S^{2}+\dots\bigg]_{2L}\\
   && +
   \bigg[\left(\frac{1}{16 \mathcal{J}^6}+\dots\right)\,\S+\left(\frac{1}{4 \mathcal{J}^7}+\dots\right)\,\S^{2}+\dots\bigg]_{HL} \nonumber \\
   \label{eq:string1}
E_{1} &=& \bigg[
-\frac{1}{2\J^{3}}\,\S+\frac{F(\alpha)}{\J^{4}}\,\S^{2}+\dots\bigg]_{1L}+\bigg[
\frac{1}{2\J^{5}}\,\S+\frac{-\frac{1}{2}\,F(\alpha)+\frac{11}{16}}{\J^{6}}\,\S^{2}+\dots\bigg]_{2L}\nonumber \\
&&+\bigg[-\frac{1}{2\J^{7}}\,\S +\dots\bigg]_{HL}+\bigg[
\frac{1}{4\,\J^{5}}\,\S-\frac{1}{6\J^{7}}\,\S^{2}+\dots\bigg]_{non\ analytic},
\ea
where 

\hspace{4.5cm} 1L = one-loop in the dual CFT

\hspace{4.5cm} {2L = two-loops in the dual CFT}

\hspace{4.5cm} {HL = higher-loops in the dual CFT}

\medskip

\noindent
and, finally,  the {\em non analytic} terms are expected to be due to the dressing phase in the ABA Bethe equations. 
Remarkably, these terms are not dependent on the geometrical parameter $\alpha$.

For later analysis, it is convenient to 
write the dressing contributions together with that of the analogous terms for the folded string in $AdS_{5}\times S^{5}$
\ba
\label{eq:dressing-ads5}
E_{1, \ AdS_{5}}^{\rm dressing} &=& 
\bigg[\frac{\left(\mathcal{J}^2+2\right) \coth ^{-1}\left(\sqrt{\mathcal{J}^2+1}+\mathcal{J}\right)}{\mathcal{J}^3
   \left(\mathcal{J}^2+1\right)^{3/2}}-\frac{1}{2 \mathcal{J}^3 \left(\mathcal{J}^2+1\right)}\bigg]\,\cS^{2}+\dots  \nonumber\\
   &=&   \bigg(\frac{0}{\J^{5}}+\frac{2}{3 \mathcal{J}^7}-\frac{16}{15 \mathcal{J}^9}
  +\dots\bigg)\,\cS^{2}+\dots,  \\ \nonumber \\
\label{eq:dressing-ads3}
E_{1, \ AdS_{3}}^{\rm dressing} &=& \frac{\coth ^{-1}\left(\sqrt{\mathcal{J}^2+1}\right)}{4 \mathcal{J}^3 \sqrt{\mathcal{J}^2+1}}\,\cS^{2}+\dots = \bigg(\frac{1}{4 \mathcal{J}^5}-\frac{1}{6 \mathcal{J}^7}+\frac{2}{15
   \mathcal{J}^9}+\dots\bigg)\,\cS^{2}+\dots.
\ea
According to the discussion in \cite{Beccaria:2012kp}, this means that non-analytic/dressing terms start
with $1/J^5 f(S/J)$   terms in  $AdS_{5}$ case and $1/J^3  f(S/J)$  terms in $AdS_{3}$ case.

\section{Algebraic curve quantization in $AdS_{3}\times S^{3}\times T^{4}$}
\label{sec:AC}

In this section we focus on the special limit case of string propagation on $AdS_{3}\times S^{3}\times T^{4}$.
This is formally obtained from the general case by taking the singular limit $\alpha\to 1$. The motivation behind the
study of this special case is that the (most interesting) dressing contributions are apparently 
independent on $\alpha$ (see (\ref{eq:string1})) and we 
we expect to clarify their origin in the simpler $T^{4}$ setup. In particular, we describe in this section the 
quantization of the classical algebraic curve relevant to $AdS_{3}\times S^{3}\times T^{4}$ by the methods
of \cite{Gromov:2008ec}. We shall show that the frequencies obtained in the previous section are precisely 
recovered by this method and we shall discuss the origin of dressing from the usual unit circle contribution
that gives the familiar Hernandez-Lopez phase in $AdS_{5}\times S^{5}$.

\subsection{Classical data}

The classical data is in terms of a two cut curve parametrized by the cut endpoints $a, b$ related to the 
classical charges and energy by the same equations as in  $AdS_{5}\times S^{5}$
\ba
\S &=& \frac{a b+1}{2\,\pi\,a b}\bigg[b\,\mathbb E\left(1-\frac{a^{2}}{b^{2}}\right)-a\,\mathbb K\left(
1-\frac{a^{2}}{b^{2}}\right)\bigg], \\
\J &=& \frac{1}{\pi\, b}\,\sqrt{(a^{2}-1)(b^{2}-1)}\,\mathbb K\left(1-\frac{a^{2}}{b^{2}}\right),\\
\mathcal E &=& \frac{a b-1}{2\,\pi\,a b}\bigg[b\,\mathbb E\left(1-\frac{a^{2}}{b^{2}}\right)+a\,\mathbb K\left(
1-\frac{a^{2}}{b^{2}}\right)\bigg].
\ea
The relevant quasi-momenta $\wt p_{1, 2, 3, 4}$ and $\wh p_{1,2,3,4}$ are the same as in $AdS_{5}\times S^{5}$ and can be written in terms of 
\ba
\wt p_{2} &=& \frac{2\,\pi\,\J\,x}{x^{2}-1}, \\
\wh p_{2} &=&  \pi-2\,\pi\,\J\,\left(\frac{a}{a^{2}-1}-\frac{x}{x^{2}-1}\right)\,
\sqrt{\frac{(a^{2}-1)(b^{2}-x^{2})}{(b^{2}-1)(a^{2}-x^{2})}}\nonumber \\
&&+\frac{8\,i\,\pi\,a\,b\,\S}{(b-a)(ab+1)}
\mathbb F\left(i\,\mbox{arcsinh}\left(\sqrt{-\frac{(a-b)(a-x)}{(a+b)(a+x)}}\right), \frac{(a+b)^{2}}{(a-b)^{2}}\right)
\nonumber \\
&&+\frac{2\,i\,\pi\,(a-b)\,\J}{\sqrt{(a^{2}-1)(b^{2}-1)}}\,\mathbb E\left(i\,\mbox{arcsinh}\left(\sqrt{-\frac{(a-b)(a-x)}{(a+b)(a+x)}}\right), \frac{(a+b)^{2}}{(a-b)^{2}}\right),
\ea
as
\be
\wt p_{1} = -\wt p_{3} = -\wt p_{4} = \wt p_{2}, \qquad 
-\wh p_{1}(1/x) = -\wh p_{3}(x) = \wh p_{4}(1/x) = \wh p_{2}(x).
\ee

\subsection{Algebraic curve one-loop quantization and comparison with world-sheet}

The relevant string polarizations can be assigned at the two sheets $A, B$ of the disconnected
$AdS_{3}$ algebraic curve (see for instance \cite{Beccaria:2012kb}) according to the following table
where we adopt the  $AdS_{5}\times S^{5}$ labeling of off-shell frequencies of \cite{Beccaria:2012mx}
\be
\begin{array}{||c|c|c||}
\hline\hline
\mbox{polarization} & \mbox{sheet(s)} & \mbox{off-shell frequency} \\
(\wt 2, \wt 3) & A+B & (1+1)\times \Omega_{S} \\
(\wh 1, \wh 4) & B & \Omega_{1} \\
(\wh 2, \wh 3) & A & \Omega_{A} \\
&& \\
\hline
&& \\ 
(\wh 2, \wt 4) & A & -2\times \Omega_{3}, \\
(\wt 2, \wh 4) & B & -2\times \Omega_{4} \\
\hline\hline
\end{array}
\ee
where the off-shell frequencies are 
\ba
\Omega_{S}(x) &=& \frac{2}{ab-1}\,\frac{\sqrt{a^{2}-1}\sqrt{b^{2}-1}}{x^{2}-1}, \\
\Omega_{A}(x) &=& \frac{2}{ab-1}\,\bigg(1-\frac{\sqrt{x-a}\sqrt{x+a}\sqrt{x-b}\sqrt{x+b}}{x^{2}-1}\bigg),\\
\Omega_{1}(x) &=& -2-\Omega_{A}(1/x), \\
\Omega_{3}(x) &=& \frac{1}{2}(\Omega_{A}(x)+\Omega_{S}(x)), \\
\Omega_{4}(x) &=& \frac{1}{2}(\Omega_{S}(x)-\Omega_{A}(1/x))-1. 
\ea
For each polarization $(I,J)$, we compute $x_{n}^{(I,J)}$ from 
\be
p^{I}(x^{(I,J)}_{n})-p^{J}(x^{(I,J)}_{n}) = 2\,\pi\,n,
\ee
and plug it into the relevant off-shell frequency. Once this is done, we find the following relations with the
world-sheet frequencies
\ba
\kappa\,\Omega_{S}(x^{(\wt 2, \wt 3)}_{n}) &=& \sqrt{n^{2}+\J^{2}}+\Delta\Omega_{S}, \\
\kappa\,\Omega_{1}(x^{(\wh 1, \wh 4)}_{n}) &=& \omega^{B\,(7)}_{n}+\Delta\Omega_{1}, \\
\kappa\,\Omega_{A}(x^{(\wh 2, \wh 3)}_{n}) &=& \omega^{B\,(8)}_{n}+\Delta\Omega_{A}, \\
\kappa\,\Omega_{3}(x^{(\wh 2, \wt 4)}_{n}) &=& \omega^{F}_{n}(1, -1)+\Delta\Omega_{3}, \\
\kappa\,\Omega_{4}(x^{(\wt 2, \wh 4)}_{n}) &=& \omega^{F}_{n}(1, 1)+\Delta\Omega_{4}, 
\ea
where the shifts $\Delta\Omega$ are collected in App.~(\ref{app:AC-shifts}).
These shifts are independent on $n$ and separately cancel on each sheet:
\ba
\Delta\Omega_{S}+\Delta\Omega_{1}-2\,\Delta\Omega_{4} &=& \mc O(\S^{3}), \\
\Delta\Omega_{S}+\Delta\Omega_{A}-2\,\Delta\Omega_{3} &=& \mc O(\S^{3}),
\ea
thus proving the complete equivalence between the world-sheet and quantized algebraic curve computation of the 
frequencies.

\subsection{Dressing from the algebraic curve}

The one-loop energy can be computed in the so-called algebraic curve regularization
by transforming the sum over over frequencies into a contour integral and deforming the contour 
along the unit circumference plus additional cut contributions (see for instance the detailed discussion in 
\cite{Gromov:2007cd}). The unit circumference contour is expected to give the dressing contribution.
Let us denote by $A$ the various string polarizations, label $(A_{1}, A_{2})$ the associated pair of quasimomenta and
 define $(-1)^{F_{A}}$ to be the Bose-Fermi sign taking into account statistics. The relevant formula reads
\be
\label{eq:contour}
E_{1}^{\rm dressing} = \frac{1}{4\,\pi}\int_{-1}^{1} dx\,\sum_{A}(-1)^{F_{A}}\,\Omega^{(A)}(x)
\, (p_{A_{1}}(x)-p_{A_{2}}(x))',
\ee
where the integral is computed along the upper unit half-circumference. Expanding in powers of $\S$, we find
\ba
\label{eq:AC-dressing}
E_{1}^{\rm dressing} &=& \S^{2}\,\int_{-1}^{1} dx \,
\frac{8 \left(4 \mathcal{J}^4 x^6+\frac{1}{4} \left(x^2-1\right)^3 \left(x^2+1\right)+\mathcal{J}^2 \left(x^6+3 x^4-3 x^2-1\right)
   x^2\right)}{\left(4 \mathcal{J}^3 x^2-\mathcal{J} \left(x^2-1\right)^2\right)^3}+ \mc O(\S^{3})\nonumber \\
   &=& \S^{2}\bigg[
   \frac{\coth ^{-1}\left(\sqrt{\mathcal{J}^2+1}\right)}{4 \mathcal{J}^3 \sqrt{\mathcal{J}^2+1}}+
   \frac{1}{2\,\J^{5}}\bigg]+\mc O(\S^{3}).
\ea
Comparing with (\ref{eq:dressing-ads3}), we see that there is an extra piece $\frac{1}{2\,\J^{5}}$. This is due to 
a regularization issue that we now explain.

\subsection{Matching of AC and  WS regularizations}
\label{sec:matching}

It is know that there can be a regularization mismatch between the standard world-sheet computation and the algebraic curve one. The reason is simple. The one-loop energy is evaluated as a sum over frequencies $\omega_{n}^{A}$
where $A$ labels the various modes. The world-sheet regularization amounts to summing over $n$ with a cut-off
$|n|\le N$ independent on $A$ and taking the finite limit $N\to \infty$. Instead, in the algebraic curve regularization, one
takes a fixed spectral radius cut-off $|x-1|\ge 1+\eps$ and sends $\eps\to 0$. This cut-off is effectively equivalent to 
different mode number cut-offs $N_{A}$ and there is a mismatch between the two procedures. The mismatch can be 
computed as follows (see \cite{Beccaria:2012kb} for the example of a circular string classical solution).

Let us have in mind the example of the short folded string, but try to be as general as possible.
Setting
\be
x = 1+\varepsilon\J+\frac{\varepsilon}{2}\,\J^{2},
\ee
we have
\ba
N_{A}(x) &=& \frac{p_{A_{1}}(x)-p_{A_{2}}(x)}{2\,\pi} = \frac{1}{\varepsilon}+\Delta^{(0)}_{A}+\varepsilon\,\Delta^{(1)}_{A}+\dots, \\
\kappa\,\omega_{A}(x) &=&\frac{1}{\varepsilon}+\Omega^{(0)}_{A}+\varepsilon\,\Omega^{(1)}_{A}+\dots,
\ea
where $A = (A_{1}, A_{2})$ are the string polarizations. Inverting the relation between $\varepsilon$ and $N_{A}$:
\be
\varepsilon = \frac{1}{N_{A}}+\frac{\Delta^{(0)}_{A}}{N_{A}^{2}}+\frac{{\Delta_{A}^{(0)}}^{2}+\Delta_{A}^{(1)}}{N_{A}^{3}}+\dots, 
\ee
we can write the large mode number expansion of each world-sheet frequency
\be
\kappa\,\Omega_{A} = N_{A}+(\Omega_{A}^{(0)}-\Delta_{A}^{(0)})+\frac{\Omega_{A}^{(1)}-\Delta_{A}^{(1)}}{N_{A}}+\dots .
\ee
Let $C_{A}$ be the numerical integers that are used to combine the various modes. We know that $\sum_{A}C_{A}N_{A}=0$, so 
\be
\sum_{A} C_{A} = \sum_{A}C_{A}\Delta^{(0)}_{A} = \sum_{A}C_{A}\Delta^{(1)}_{A} = 0.
\ee
Also, UV convergence requires
\be
\sum_{A}C_{A}(\Omega^{(0)}_{A}-\Delta^{(0)}_{A}) = \sum_{A}C_{A}(\Omega^{(1)}_{A}-\Delta^{(1)}_{A})=0,
\ee
or, due to the previous relation
\be
\sum_{A}C_{A}\Omega^{(0)}_{A} = \sum_{A}C_{A}\Omega^{(1)}_{A}=0.
\ee
These relations actually hold for each of the two sheets separately.
The mismatch due to regularization is \cite{Beccaria:2012kb}
\ba
\kappa\,\delta^{\rm AC-WS} &=& \lim_{\varepsilon\to 0}\sum_{A}C_{A}\bigg[
\frac{1}{2}N^{2}+(\Omega^{(0)}_{A}-\Delta^{(0)}_{A})\,N
\bigg]^{\frac{1}{\varepsilon}+\Delta^{(0)}_{A}+\varepsilon \Delta^{(1)}_{A}+\dots}_{\frac{1}{\varepsilon}} = \nonumber \\
&=& \sum_{A} C_{A}\bigg[\Delta_{A}^{(1)}-\frac{1}{2}{\Delta^{(0)}_{A}}^{2}+\Omega^{(0)}_{A}\Delta^{(0)}_{A}\bigg],
\ea
and, using again the sum rules
\be
\delta^{\rm AC-WS} =\frac{1}{\kappa} \sum_{A} C_{A}\bigg[-\frac{1}{2}{\Delta^{(0)}_{A}}^{2}+\Omega^{(0)}_{A}\Delta^{(0)}_{A}\bigg].
\ee
For the folded string in the short limit we report the values of $\Delta^{(0)}$ and $\Omega^{(0)}$ in 
App.~(\ref{app:AC-details})). We find
\be
\delta^{\rm AC-WS} = \frac{\S^{2}}{2\,\J^{5}}+\mc O(\S^{3}),
\ee
explaining the extra term in (\ref{eq:AC-dressing}).


\section{Large $J$ limit of the ABA Bethe equations}
\label{sec:Bethe}

In \cite{OhlssonSax:2011ms}, a set of all-loop asymptotic Bethe equations (ABA) has been proposed 
to describe strings on $AdS_{3}\times S^{3}\times S^{3}\times S^{1}$, with symmetry $d(2,1;\alpha)^{2}$, 
valid for all values of $\alpha$. In the spirit of \cite{Beccaria:2012kp}, it is interesting to compare the large $J$ and fixed $S$ expansion of the Bethe Anstaz energy at weak coupling one and two-loops level with the large $\J$ expansion of the classical and one-loop string energy. The aim of the comparison is that of checking whether weak-coupling contributions of the form 
$\frac{S^{p}}{J^{q}}\,\lambda^{n}$ match the analogous terms $\frac{\S^{p}}{\J^{q}}\,\lambda^{n+\frac{p-q}{2}}$
in the string theory. Such a matching would show that these terms are protected from possible non-trivial effects that 
could appear in the extrapolation between weak and strong coupling. 

In particular, one can expect that 
these non-trivial effects are related to the dressing phase(s) in the Bethe equations and do not modify the above contributions at least for low integer values of $n$, {\em i.e.} one and two loops contributions have a chance of being equal to the corresponding terms in the string theory. For these reasons, we shall first analyze the ABA equations without 
introducing any dressing phase(s) and in a second step shall discuss the role of dressing.

\subsection{ABA equations without dressing}

The relevant subset of Bethe equations involves roots at two coupled nodes. In the notation of \cite{OhlssonSax:2011ms} the logarithmic form of the equations suitable for the analysis of the folded string solution can be written
\ba 
J\,\log\bigg(\frac{x^{+}_{1, i}}{x^{-}_{1,i}}\bigg) &=& \sum_{k\neq i}^{S}\log\bigg(
\frac{1-\frac{h}{x^{+}_{1,i}\,x^{-}_{1,k}}}{1-\frac{h}{x^{-}_{1,i}\,x^{+}_{1,k}}}\,\sigma^{2}_{1}(x_{1,i},x_{1,k})\bigg)\,+
\sum_{k=1}^{S}\log\bigg(
\frac{x^{-}_{1,i}-x^{+}_{3,k}}{x^{+}_{1,i}-x^{-}_{3,k}}\bigg)+2\,\pi\,i\,n_{i}, \nonumber \\
&& \\
J\,\log\bigg(\frac{x^{+}_{3, i}}{x^{-}_{3,i}}\bigg) &=& \sum_{k\neq i}^{S}
\log\bigg(\frac{1-\frac{h}{x^{+}_{3,i}\,x^{-}_{3,k}}}{1-\frac{h}{x^{-}_{3,i}\,x^{+}_{3,k}}}\,\sigma^{2}_{3}(x_{3,i},x_{3,k})\bigg)+
\sum_{k=1}^{S}\log\bigg(
\frac{x^{-}_{3,i}-x^{+}_{1,k}}{x^{+}_{3,i}-x^{-}_{1,k}}\bigg).
\ea
Here, the quantities $x_{\ell,i}^{\pm}$, $\ell=1,3$, $i=1, \dots, S$,  are defined on the nodes 1, 3 as
\be
x_{1, i}^{\pm} = x\bigg(u_{1, i}\pm i\,\alpha\bigg), \qquad
x_{3, i}^{\pm} = x\bigg(u_{3, i}\pm i\,(1-\alpha)\bigg),
\ee
where
\be
x(u) = \frac{u}{2}\,\bigg(1+\sqrt{1-\frac{4\,h}{u^{2}}}\bigg).
\ee
The coupling $h(\lambda)$ is an interpolating coupling with the  strong coupling expansion at $\lambda\gg 1$
$h(\lambda) = \frac{\sqrt\lambda}{2\,\pi}+\mc O(1)$.
The energy and momentum are computed from 
\ba
\label{eq:energy}
E(S, J) &=& i\,h\,\sum_{\ell=1,3}\sum_{k=1}^{S}\bigg(\frac{1}{x^{+}_{\ell,k}}
-\frac{1}{x^{-}_{\ell,k}}\bigg), \\
e^{i\,P} &=& \prod_{\ell=1,3}\prod_{k}\frac{x^{+}_{\ell,k}}{x^{-}_{\ell,k}} = 1.
\ea
We shall write the weak-coupling expansion of the energy as 
\be
E(S,J) = 4\,\pi^{2}\,h\,E^{(1)}(S,J)+(4\,\pi^{2}\,h)^{2}\,E^{(2)}(S,J)+\dots, 
\ee
where the choice of normalization of $E^{(n)}$ will be explained later.

\subsubsection{Large $J$ expansion of the one-loop energy}

The one-loop Bethe roots $u_{\ell, i}$ of the folded string are symmetric under $u\to -u$ 
\ba
u_{1,i} &=& (U_{1}, U_{2}, \dots, U_{\frac{S}{2}}, -U_{1}, -U_{2}, \dots, -U_{\frac{S}{2}}), \\
u_{3,i} &=& (U_{\frac{S}{2}+1}, U_{\frac{S}{2}+2}, \dots, U_{S}, -U_{\frac{S}{2}+1}, -U_{\frac{S}{2}+2}, \dots, -U_{S}),
\ea
and the mode numbers are~\footnote{Notice that the mode numbers of the 3-roots are trivial with the standard branch of the logarithm.}
\be
n_{i} = (-1, \dots, -1, 1, \dots, 1).
\ee
As we said, we first consider the ABA without dressing and put $\sigma_{1,3}\to 1$.

The numerical analysis of the large $J$ Bethe roots suggests the following Ansatz\footnote{A posteriori, the expansion based on this Ansatz will match perfectly a numerical fit to the various $1/J$ coefficients of the numerical Bethe roots.}
\ba
\label{eq:AnsatzU1}
u_{i} &=& -\frac{J}{\pi}+c^{(1)}_{i}\,\sqrt J+c^{(2)}_{i}+c^{(3)}_{i}\,\frac{1}{\sqrt J}
+c^{(4)}_{i}\,\frac{1}{J}+\cdots, \qquad i=1, \dots, \frac{S}{2}, \\
\label{eq:AnsatzU3}
u_{i} &=& -\frac{J}{\pi}+c^{(1)}_{i-\frac{S}{2}}\,\sqrt J+c^{(2)}_{i}+c^{(3)}_{i}\,\frac{1}{\sqrt J}
+c^{(4)}_{i}\,\frac{1}{J}+\cdots, \qquad i=\frac{S}{2}+1, \dots, S.
\ea
The correlation between the coefficients $c^{(1)}_{i}$ is a non trivial remark. Expanding the Bethe equations
we find the following remarkable relation 
\be
\pi^{2}\,c^{(1)}_{i} = 2\,\sum_{j\neq i}\frac{1}{c^{(1)}_{i}-c^{(1)}_{j}}, \qquad i=1, \dots, \frac{S}{2}.
\ee
It implies that the coefficients $c_{i}^{(1)}$ for $i=1, \dots, \frac{S}{2}$ are the roots of 
\be
H_{\frac{S}{2}}\bigg(\frac{\pi}{\sqrt 2}\,c^{(1)}\bigg)=0,
\ee
where $H_{n}$ is the n-th Hermite polynomial. 
The other coefficients obey some analytical relations like (here again $i=1, \dots, \frac{S}{2}$)
\ba
c^{(2)}_{i+\frac{S}{2}} &=& c^{(2)}_{i}-\cot(\pi \alpha), \\
c^{(3)}_{i+\frac{S}{2}} &=& c^{(3)}_{i}+\frac{\pi^{2}(2\alpha-1)}{2\sin^{2}(\pi \alpha)}\,c^{(1)}_{i}, 
\ea
but basically must be determined numerically keeping $\alpha$ generic. Plugging the resulting expansion of the Bethe roots in the expression for the energy (\ref{eq:energy}) we obtain the following expression for the first three orders of the 
large $J$ expansion of the one-loop energy $E^{(1)}$ 
\be
\label{eq:gaugeresult1}
E^{(1)}(S,J) = \frac{S}{2\,J^{2}}-\bigg(\frac{S^{2}}{4}+\frac{S}{2}\bigg)\,\frac{1}{J^{3}}+
\bigg[
\frac{3}{16}\,S^{3}+F(\alpha)\,S^{2}+\frac{S}{2}
\bigg]\,\frac{1}{J^{4}}+\dots,
\ee
where $F(\alpha)$ has been defined in (\ref{eq:alpha}).

\subsection{Large $J$ expansion at two-loops}

Expanding at order $\mathcal O(h^{2})$ the Bethe roots are written as
\be
u_{i} = u_{i}^{(0)}+h\,u_{i}^{(1)}+\dots,
\ee
where $u_{i}^{(0)}$ are given in (\ref{eq:AnsatzU1}) and (\ref{eq:AnsatzU3}) and 
\ba
u_{i}^{(1)} &=& -\frac{2\,\pi}{J}+\frac{d^{(1)}_{i}}{J^{3/2}}+\frac{d^{(2)}_{i}}{J^{2}}+
\frac{d^{(3)}_{i}}{J^{5/2}}+\frac{d^{(4)}_{i}}{J^{3}}+\dots, \qquad i=1, \dots, \frac{S}{2}, \\
u_{i}^{(1)} &=& -\frac{2\,\pi}{J}+\frac{d^{(1)}_{i-\frac{S}{2}}}{J^{3/2}}+\frac{d^{(2)}_{i-\frac{S}{2}}}{J^{2}}+
\frac{d^{(3)}_{i}}{J^{5/2}}+\frac{d^{(4)}_{i}}{J^{3}}+\dots, \qquad i=\frac{S}{2}+1, \dots, S.
\ea
Again, the correlation between the coefficients $d^{(1)}_{i}$ and $d^{(2)}_{i}$ is a non trivial remark.
Evaluating the two-loop energy we find
\be
\label{eq:gaugeresult2}
E^{(2)} = -\frac{S}{8\,J^{4}}+\bigg(-\frac{S^{2}}{4}+\frac{S}{2}\bigg)\,\frac{1}{J^{5}}+
\bigg[\frac{11}{32}\,S^{3}+\bigg(-\frac{1}{2}\,F(\alpha)+\frac{11}{16}\bigg)\,S^{2}-\frac{11}{8}\,S
\bigg]\,\frac{1}{J^{6}}+\dots.
\ee

\subsubsection{Comparison with string theory}

If we assume the following weak-coupling expansion of the interpolating coupling $h(\lambda)$
\be
h(\lambda) = \frac{\lambda}{4\,\pi^{2}}+\mc O(\lambda^{2}), 
\ee
then, our one and two-loop results (\ref{eq:gaugeresult1}, \ref{eq:gaugeresult2}) nicely reproduce the 
{\em 1L} and {\em 2L} terms in (\ref{eq:string0}, \ref{eq:string1}). The {\em HL} terms would require a three or higher-loop computation at weak-coupling but are expected to violate matching being unprotected. The most interesting point of
the comparison is the fact that the {\em non analytic} terms in (\ref{eq:string1}) are clearly absent. They require the introduction of 
dressing phases to which we devote the next section.

\section{ABA with dressing at $\alpha=1$}
\label{sec:Dressing}

We discuss dressing in the ABA for the $\alpha=1$ case of string on $AdS_{3}\times S^{3}\times T^{4}$
due to several reasons
\begin{enumerate}
\item[a)] We already observed that dressing effects computed by world-sheet methods are independent on $\alpha$, so this should be a simplification that does not spoil any essential feature.
\item[b)] We want to build on the algebraic curve detailed description that we have presented in this case.
\item[c)] We want to make contact with the discussion of semiclassical dressing discussed in \cite{Beccaria:2012kb}.
\end{enumerate}

\subsection{Prediction according to the BLMT proposal}

The relevant ABA has been proposed in \cite{Babichenko:2009dk} and its reduction to the $\mathfrak{sl}(2)$ sector
gives the same Bethe equations as in $AdS_{5}\times S^{5}$ although with a possibly different phase. They read
\be
\label{eq:sl2-bethe}
\bigg(\frac{x^{+}_{i}}{x^{-}_{i}}\bigg)^{J} = \prod_{j\neq i}^{S}
\frac{x_{i}^{-}-x_{j}^{+}}{x_{i}^{+}-x_{j}^{-}}\,
\frac{1-\frac{1}{x_{i}^{+}\,x_{j}^{-}}}{1-\frac{1}{x_{i}^{-}\,x_{j}^{+}}}\,e^{2\,i\,\theta_{ij}},
\ee
where the phase is 
\be
\theta_{ij} = \sum_{r,s \ge 1} \bigg[h\,c_{r,s}^{(0)}+c_{r,s}^{(1)}+\dots\bigg]\,\,q_{s}(u_{i})\,q_{r}(u_{j}).
\ee
in terms of the higher charges
\be
q_{r} = \frac{i}{r-1}\,\bigg(\frac{1}{(x^{+})^{r-1}}-\frac{1}{(x^{-})^{r-1}}\bigg).
\ee
Notice that for these $\alpha=1$ Bethe equations the map $x(u)$ is basically the same as in $AdS_{5}\times S^{5}$
\be
\label{eq:babichenko-notation}
x+\frac{1}{x} = \frac{u}{h},\qquad 
x^{\pm}+\frac{1}{x^{\pm}} = \frac{u\pm \frac{i}{2}}{h}.
\ee
and the energy has a factor 2 fixed by the $\alpha=1$ dispersion relations
\be
E_{1} = 2\,i\,h\,\sum_{i=1}^{S}\bigg(\frac{1}{x^{+}_{i}}-\frac{1}{x^{-}_{i}}\bigg).
\ee
We now expand $E_{1}$ at large $J$ and extract the dressing contribution which is $\cS^{2}$ times 
a series with odd powers of $1/\J$. To do so, we have to fix the relation between $\sqrt\lambda$ and $h$ at strong coupling that we take equal to the $AdS_{5}\times S^{5}$ one, i.e. $h = \frac{\sqrt\lambda}{4\,\pi}+\dots$.
Then, we find
\ba
E_{1}^{\rm dressing} &=& \bigg(
\frac{a_{12}}{4\,\J^{5}}+\frac{-3\, a_{12}+a_{14}-a_{23}}{16\,\J^{7}} \nonumber\\
&&
\qquad +\frac{-10\, a_{12}+5\,a_{14}-a_{16}+5\, a_{23}-a_{25}+a_{34}}{64\,\J^{9}}+\dots
\bigg)\,\cS^{2}+\dots,
\ea
where
\be
a_{r,s}=c_{r,s}^{(1)}-c_{s,r}^{(1)}.
\ee
If we plug in this expression the Hernandez-Lopez values \cite{Hernandez:2006tk} valid for the $AdS_{5}\times S^{5}$ case
\be
c_{r,s}^{(1)} = -8\,\frac{1-(-1)^{r+s}}{2}\,\frac{(r-1)(s-1)}{(r+s-2)(s-r)},
\ee
we recover the expansion in (\ref{eq:dressing-ads5}). If instead, we use the BLMT coefficients that have been proposed
in  \cite{Beccaria:2012kb}, i.e.
\be
c_{r,s}^{(1)} = 2\,\frac{1-(-1)^{r+s}}{2}\,\frac{s-r}{r+s-2},
\ee
we find
\ba
\label{eq:dressing-discrete}
E^{\rm dressing}_{1} &=& \bigg(\frac{1}{\J^{5}}-\frac{7}{12\,\J^{7}}+\frac{109}{240\,\J^{9}}+\dots\bigg)\,\S^{2}+\mc O(\S^{3})\nonumber \\
&&  = \bigg[
\frac{\coth ^{-1}\left(\sqrt{\mathcal{J}^2+1}\right)}{2 \mathcal{J}^3 \sqrt{\mathcal{J}^2+1}}
+\frac{1}{2\,\J^{4}\,\sqrt{\J^{2}+1}}\bigg]\,\cS^{2}+\mc O(\S^{3}).
\ea
The analytic expression resumming the three terms of the large $\J$ series is not a mere conjecture. Indeed, 
it will be strongly motivated and explained in the next section devoted to a discussion.

\subsection{Discussion}

The classical scaling limit of (\ref{eq:sl2-bethe}) can be written as
\be
2\,\pi\,n +\frac{4\,\pi\,\J\,x}{x^{2}-1} = 2\,H(x)-2\,
\frac{G(0)}{x^{2}-1},
\ee 
where the functions $H$ and $G$ are defined as 
\be
G(x) = \sum_{k=1}^{S} \frac{\wh\alpha(x_{k})}{x-x_{k}},\qquad
H(x) = \sum_{k=1}^{S} \frac{\wh\alpha(x)}{x-x_{k}},\qquad\qquad \wh\alpha(x) = \frac{1}{h}\,\frac{x^{2}}{x^{2}-1}.
\ee
The analysis of \cite{Beccaria:2012kb} shows that one-loop semiclassical effects associated with the 
unit circumference in the spectral plane are incorporated by adding to the r.h.s. of this equation the potential term 
\be
\mc V(x) =\int_{-1}^{1}\frac{dy}{2\,\pi}\,\Big[\partial_{y}\,G(y)\,\frac{\wh \alpha(x)}{x-y}
+\partial_{y}\, G(1/y)\,\frac{\wh \alpha(1/x)}{1/x-y}\Big],
\ee
where the notation is $\int_{-1}^{1} = \frac{1}{2}\int_{C^{+}}+\frac{1}{2}\int_{C^{-}}$ and the half circumferences
$C^{\pm}$ (and their orientation) are defined in the caption of figure 4 of \cite{Gromov:2007cd}. This term cannot be 
interpreted as a phase. This is possible up to a remainder if we integrate by parts the second term in the integral
\be
\mc V(x) = \mc V_{\rm phase}(x)+\Delta \mc V(x),
\ee
where
\ba
\mathcal{V}_{\rm phase}(x) &=&  \int_{-1}^{1}\frac{dy}{2\,\pi}\Big[G'(y)\,\frac{\wh \alpha(x)}{x-y}
-G(1/y)\,\Big(\frac{\wh \alpha(1/x)}{1/x-y}\Big)'\Big], \\
\Delta \mathcal V(x) &=& \frac{\wh \alpha(1/x)}{2\,\pi}\,\Big[
\frac{G(1)}{1/x-1}-\frac{G(-1)}{1/x+1}\Big].
\ea
The role of $\Delta \mc V$ is unclear at the moment. A possible interpretation of this term is that of a modification
that it is necessary to introduce at the level of the discrete Bethe equations (\ref{eq:sl2-bethe}) going beyond the 
recipes described in \cite{Zarembo:2010yz} and necessary in order to match semiclassical string theory at one-loop.
Hints at such non-trivial modifications are also present in the recent analyses in \cite{Rughoonauth:2012qd} and 
\cite{Borsato:2012ss}.

What can be remarked here, is that the phase term $\mc V_{\rm phase}(x)$ is fully consistent with the 
result (\ref{eq:dressing-discrete}). Indeed, the integration by parts (and neglecting the controversial term $\Delta\mc V$)
amounts to make the same transformation in (\ref{eq:contour}). In the notation of Sec.~(\ref{sec:matching}), 
integration by parts on a subset $\mathcal A$ of polarizations gives the contribution
\be
\mbox{IBP}_{\mathcal A} = \lim_{x\to 1} \sum_{A\in \mathcal A} C_{A}\omega_{A}(x) N_{A}(x)
= \frac{1}{\kappa}\,\sum_{A\in \mathcal A} C_{A}\bigg[
\Delta^{(1)}_{A}+\Omega^{(1)}_{A}+\Omega^{(0)}_{A}\Delta^{(0)}_{A}
\bigg].
\ee
If $\mathcal A$ is the second sheet, then we simply have
\be
\mbox{IBP}_{\mbox{\scriptsize second sheet}} = \frac{1}{\kappa}\,\sum_{A\in \mbox{\scriptsize second sheet}} C_{A} \Omega^{(0)}_{A}\Delta^{(0)}_{A}.
\qquad
\ee
Evaluating this quantity according to the results in App.~(\ref{app:AC-details}), we find 
\be
\mbox{IBP}_{\mbox{\scriptsize second sheet}} = \bigg(
\frac{1}{2\,\J^{5}}-\frac{1}{2\,\J^{4}\,\sqrt{1+\J^{2}}}
\bigg)\,\S^{2}+\mc O(\S^{3}).
\ee
This is precisely the piece that we have to add to (\ref{eq:dressing-discrete}) in order to recover the 
old AC regularized result equivalent, up to regularization correction, to the world-sheet computation~\footnote{
To be more precise, that there is also a global factor of 2 compared to the  $0<\alpha<1$. This is known to be related to the 
extra massless modes that are present at $\alpha=1$ as discussed in  \cite{Beccaria:2012kb}
for a circular string solution as well as for the long folded string.}.

\subsection*{Acknowledgments }

We would like to thank A. A. Tseytlin and  F. Levkovich-Maslyuk for very useful and pleasant discussions. 
 

\appendix

\numberwithin{equation}{section}
\section{Simplification of the quadratic fermionic fluctuation operator}
\label{app:fermionic}

In this appendix, we explain how to deal with the fermionic operator (\ref{eq:fermionic}). It is convenient to set 
 \be
 f_{1}(\sigma) = \frac{\kappa\,w\,\J}{2\,(\rho'^{2}+\J^{2})}, \quad
 f_{2}(\sigma) = \frac{\rho'}{2}, \quad
 f_{3}(\sigma) = \frac{\sqrt{\rho'^{2}+\J^{2}}-\rho'}{2}.
 \ee
 Squaring $\mathcal D_{F}$ we find
 \ba
-\mathcal D_{F}^{2} &=& -\partial^{2}+f_{1}^{2}+2\,\alpha\,(\alpha-1)\,f_{3}\,(2\,f_{2}+f_{3})
 +\Gamma_{02}\,\bigg[f_{2}'+2\,f_{2}\,\partial_{\sigma}+f_{3}'+2\,f_{3}\,\partial_{\sigma}
\bigg]\nonumber  \\
&& +\Gamma_{12}\,\bigg[-2\,(f_{2}+f_{3})\,\partial_{\tau}
\bigg]+\Gamma_{25}\,\bigg[\sqrt\alpha\,(f_{1}'+2\,f_{1}\,\partial_{\sigma})
\bigg]+\Gamma_{28}\,\bigg[\sqrt{1-\alpha}\,(f_{1}'+2\,f_{1}\,\partial_{\sigma})
\bigg]\nonumber  \\
&& +\Gamma_{1234}\,\bigg[-2\,\alpha\,f_{1}\,(f_{2}+f_{3})
\bigg]+\Gamma_{1267}\,\bigg[-2\,(1-\alpha)\,f_{1}\,(f_{2}+f_{3})
\bigg]\nonumber  \\
&& +\Gamma_{1345}\,\bigg[\sqrt\alpha\,(f_{2}'+\alpha\,f_{3}')
\bigg]+\Gamma_{1348}\,\bigg[\alpha\,\sqrt{1-\alpha}\,f_{3}'
\bigg] +\Gamma_{1567}\,\bigg[(1-\alpha)\,\sqrt\alpha\,f_{3}'
\bigg]\nonumber  \\
&& +\Gamma_{1678}\,\bigg[\sqrt{1-\alpha}(f_{2}'+(1-\alpha)\,f_{3}'
\bigg]+\Gamma_{3467}\,\bigg[-2\,\alpha\,(1-\alpha)\,f_{3}\,(2\,f_{2}+f_{3})
\bigg]\nonumber .
 \ea
At this point it is convenient to use an explicit representation of the 10d $\Gamma$ matrices. We start from the Pauli matrices:
\begin{eqnarray}
\sigma_1=
\left(
\begin{array}{cc}
 0 & 1 \\
 1 & 0 \\
\end{array}
\right),
\qquad
\sigma_2=
\left(
\begin{array}{cc}
 0 & -i \\
 i & 0 \\
\end{array}
\right),
\qquad
\sigma_3=
\left(
\begin{array}{cc}
 1 & 0 \\
 0 & -1 \\
\end{array}
\right)
\quad
\textrm{and}\quad
\epsilon = i \sigma_2.
\end{eqnarray}
and build the Spin(8) Clifford algebra as
\begin{eqnarray}
 \gamma^1=\epsilon \otimes \epsilon \otimes \epsilon & & \gamma^2=       1 \otimes \sigma_1 \otimes \epsilon \nonumber\\
 \gamma^3=   1     \otimes \sigma_3 \otimes \epsilon & & \gamma^4=\sigma_1 \otimes \epsilon \otimes        1 \nonumber\\
 \gamma^5=\sigma_3 \otimes \epsilon \otimes 1        & & \gamma^6=\epsilon \otimes    1     \otimes \sigma_1 \nonumber\\
 \gamma^7=\epsilon \otimes    1     \otimes \sigma_3 & & \gamma^8=  1 \otimes 1 \otimes 1 
\end{eqnarray}
Then, we define the $16\times 16$ matrices
\begin{eqnarray}
&&\gamma_{16}^A = 
\left(
\begin{array}{cc}
 0 & \gamma^A \\
 (\gamma^A)^T & 0 \\
\end{array}
\right)
\qquad \gamma^0_{16} = 1_{16} \qquad \gamma^9_{16} = \gamma^{12345678}_{16}\nonumber\\
&&\phantom{}\nonumber\\
&&\gamma^{\mu} = \left\{ 1,\gamma^A, \gamma^9     \right\}_{16} \qquad \bar{\gamma}^{\mu} = \left\{ - 1,\gamma^A, \gamma^9     \right\}_{16}
\end{eqnarray}
and finally the $32 \times 32$ 10d Dirac matrices according to 
\begin{eqnarray}
\Gamma_{\{\mu\}} = 
\left(
\begin{array}{cc}
 0 & \gamma^\mu \\
\bar{\gamma}^\mu & 0 \\
\end{array}
\right).
\end{eqnarray}
Our main observation is that we can find a relatively simple invertible matrix $U$ such that 
\be
U\,\Gamma_{ABC\dots}\,U^{-1},
\ee
is composed of eight $4\times 4$ blocks on the diagonal for all 11 Gamma matrix structures appearing in 
$\mathcal D_{F}^{2}$. This matrix is obtained from the eigenvectors of the commuting subset of the 11 structures and 
reads
\begin{eqnarray}
U = 
\left(
\begin{array}{cc}
 \mathcal{U}_{1,1} & \mathcal{U}_{1,2} \\
 \mathcal{U}_{2,1} & \mathcal{U}_{2,2} \\
\end{array}
\right)
\end{eqnarray}
where
{\scriptsize
\begin{eqnarray}
\mathcal{U}_{1,1}=
\left(
\begin{array}{cccccccccccccccc}
 0 & 0 & 0 & 0 & 0 & 0 & 0 & 0 & 0 & 0 & 0 & 0 & 0 & 0 & 0 & 0 \\
 0 & 0 & 0 & 0 & 0 & 0 & 0 & 0 & 0 & 0 & 0 & 0 & 0 & 0 & 0 & 0 \\
 0 & 0 & 0 & 0 & 0 & 0 & 0 & 0 & 0 & 0 & 0 & 0 & 0 & 0 & 0 & 0 \\
 0 & 0 & 0 & 0 & 0 & 0 & 0 & 0 & 0 & 0 & 0 & 0 & 0 & 0 & 0 & 0 \\
 0 & 0 & 0 & 0 & 0 & 0 & 0 & 0 & -1 & -i & -i & 1 & -i & 1 & -1 & -i \\
 -1 & -i & -i & 1 & -i & 1 & -1 & -i & 0 & 0 & 0 & 0 & 0 & 0 & 0 & 0 \\
 0 & 0 & 0 & 0 & 0 & 0 & 0 & 0 & 1 & i & -i & 1 & -i & 1 & 1 & i \\
 1 & i & -i & 1 & -i & 1 & 1 & i & 0 & 0 & 0 & 0 & 0 & 0 & 0 & 0 \\
 0 & 0 & 0 & 0 & 0 & 0 & 0 & 0 & 0 & 0 & 0 & 0 & 0 & 0 & 0 & 0 \\
 0 & 0 & 0 & 0 & 0 & 0 & 0 & 0 & 0 & 0 & 0 & 0 & 0 & 0 & 0 & 0 \\
 0 & 0 & 0 & 0 & 0 & 0 & 0 & 0 & 0 & 0 & 0 & 0 & 0 & 0 & 0 & 0 \\
 0 & 0 & 0 & 0 & 0 & 0 & 0 & 0 & 0 & 0 & 0 & 0 & 0 & 0 & 0 & 0 \\
 0 & 0 & 0 & 0 & 0 & 0 & 0 & 0 & -1 & -i & i & -1 & -i & 1 & 1 & i \\
 -1 & -i & i & -1 & -i & 1 & 1 & i & 0 & 0 & 0 & 0 & 0 & 0 & 0 & 0 \\
 0 & 0 & 0 & 0 & 0 & 0 & 0 & 0 & 1 & i & i & -1 & -i & 1 & -1 & -i \\
 1 & i & i & -1 & -i & 1 & -1 & -i & 0 & 0 & 0 & 0 & 0 & 0 & 0 & 0 \\
\end{array}
\right)\\
\mathcal{U}_{1,2}=
\left(
\begin{array}{cccccccccccccccc}
 0 & 0 & 0 & 0 & 0 & 0 & 0 & 0 & -1 & -i & -i & 1 & -i & 1 & -1 & -i \\
 -1 & -i & -i & 1 & -i & 1 & -1 & -i & 0 & 0 & 0 & 0 & 0 & 0 & 0 & 0 \\
 0 & 0 & 0 & 0 & 0 & 0 & 0 & 0 & 1 & i & -i & 1 & -i & 1 & 1 & i \\
 1 & i & -i & 1 & -i & 1 & 1 & i & 0 & 0 & 0 & 0 & 0 & 0 & 0 & 0 \\
 0 & 0 & 0 & 0 & 0 & 0 & 0 & 0 & 0 & 0 & 0 & 0 & 0 & 0 & 0 & 0 \\
 0 & 0 & 0 & 0 & 0 & 0 & 0 & 0 & 0 & 0 & 0 & 0 & 0 & 0 & 0 & 0 \\
 0 & 0 & 0 & 0 & 0 & 0 & 0 & 0 & 0 & 0 & 0 & 0 & 0 & 0 & 0 & 0 \\
 0 & 0 & 0 & 0 & 0 & 0 & 0 & 0 & 0 & 0 & 0 & 0 & 0 & 0 & 0 & 0 \\
 0 & 0 & 0 & 0 & 0 & 0 & 0 & 0 & -1 & -i & i & -1 & -i & 1 & 1 & i \\
 -1 & -i & i & -1 & -i & 1 & 1 & i & 0 & 0 & 0 & 0 & 0 & 0 & 0 & 0 \\
 0 & 0 & 0 & 0 & 0 & 0 & 0 & 0 & 1 & i & i & -1 & -i & 1 & -1 & -i \\
 1 & i & i & -1 & -i & 1 & -1 & -i & 0 & 0 & 0 & 0 & 0 & 0 & 0 & 0 \\
 0 & 0 & 0 & 0 & 0 & 0 & 0 & 0 & 0 & 0 & 0 & 0 & 0 & 0 & 0 & 0 \\
 0 & 0 & 0 & 0 & 0 & 0 & 0 & 0 & 0 & 0 & 0 & 0 & 0 & 0 & 0 & 0 \\
 0 & 0 & 0 & 0 & 0 & 0 & 0 & 0 & 0 & 0 & 0 & 0 & 0 & 0 & 0 & 0 \\
 0 & 0 & 0 & 0 & 0 & 0 & 0 & 0 & 0 & 0 & 0 & 0 & 0 & 0 & 0 & 0 \\
\end{array}
\right)
\end{eqnarray}
\begin{eqnarray}
\mathcal{U}_{1,2}=
\left(
\begin{array}{cccccccccccccccc}
 0 & 0 & 0 & 0 & 0 & 0 & 0 & 0 & 0 & 0 & 0 & 0 & 0 & 0 & 0 & 0 \\
 0 & 0 & 0 & 0 & 0 & 0 & 0 & 0 & 0 & 0 & 0 & 0 & 0 & 0 & 0 & 0 \\
 0 & 0 & 0 & 0 & 0 & 0 & 0 & 0 & 0 & 0 & 0 & 0 & 0 & 0 & 0 & 0 \\
 0 & 0 & 0 & 0 & 0 & 0 & 0 & 0 & 0 & 0 & 0 & 0 & 0 & 0 & 0 & 0 \\
 0 & 0 & 0 & 0 & 0 & 0 & 0 & 0 & 1 & -i & -i & -1 & i & 1 & -1 & i \\
 1 & -i & -i & -1 & i & 1 & -1 & i & 0 & 0 & 0 & 0 & 0 & 0 & 0 & 0 \\
 0 & 0 & 0 & 0 & 0 & 0 & 0 & 0 & -1 & i & -i & -1 & i & 1 & 1 & -i \\
 -1 & i & -i & -1 & i & 1 & 1 & -i & 0 & 0 & 0 & 0 & 0 & 0 & 0 & 0 \\
 0 & 0 & 0 & 0 & 0 & 0 & 0 & 0 & 0 & 0 & 0 & 0 & 0 & 0 & 0 & 0 \\
 0 & 0 & 0 & 0 & 0 & 0 & 0 & 0 & 0 & 0 & 0 & 0 & 0 & 0 & 0 & 0 \\
 0 & 0 & 0 & 0 & 0 & 0 & 0 & 0 & 0 & 0 & 0 & 0 & 0 & 0 & 0 & 0 \\
 0 & 0 & 0 & 0 & 0 & 0 & 0 & 0 & 0 & 0 & 0 & 0 & 0 & 0 & 0 & 0 \\
 0 & 0 & 0 & 0 & 0 & 0 & 0 & 0 & 1 & -i & i & 1 & i & 1 & 1 & -i \\
 1 & -i & i & 1 & i & 1 & 1 & -i & 0 & 0 & 0 & 0 & 0 & 0 & 0 & 0 \\
 0 & 0 & 0 & 0 & 0 & 0 & 0 & 0 & -1 & i & i & 1 & i & 1 & -1 & i \\
 -1 & i & i & 1 & i & 1 & -1 & i & 0 & 0 & 0 & 0 & 0 & 0 & 0 & 0 \\
\end{array}
\right)
\\
\mathcal{U}_{2,2}=
\left(
\begin{array}{cccccccccccccccc}
 0 & 0 & 0 & 0 & 0 & 0 & 0 & 0 & 1 & -i & -i & -1 & i & 1 & -1 & i \\
 1 & -i & -i & -1 & i & 1 & -1 & i & 0 & 0 & 0 & 0 & 0 & 0 & 0 & 0 \\
 0 & 0 & 0 & 0 & 0 & 0 & 0 & 0 & -1 & i & -i & -1 & i & 1 & 1 & -i \\
 -1 & i & -i & -1 & i & 1 & 1 & -i & 0 & 0 & 0 & 0 & 0 & 0 & 0 & 0 \\
 0 & 0 & 0 & 0 & 0 & 0 & 0 & 0 & 0 & 0 & 0 & 0 & 0 & 0 & 0 & 0 \\
 0 & 0 & 0 & 0 & 0 & 0 & 0 & 0 & 0 & 0 & 0 & 0 & 0 & 0 & 0 & 0 \\
 0 & 0 & 0 & 0 & 0 & 0 & 0 & 0 & 0 & 0 & 0 & 0 & 0 & 0 & 0 & 0 \\
 0 & 0 & 0 & 0 & 0 & 0 & 0 & 0 & 0 & 0 & 0 & 0 & 0 & 0 & 0 & 0 \\
 0 & 0 & 0 & 0 & 0 & 0 & 0 & 0 & 1 & -i & i & 1 & i & 1 & 1 & -i \\
 1 & -i & i & 1 & i & 1 & 1 & -i & 0 & 0 & 0 & 0 & 0 & 0 & 0 & 0 \\
 0 & 0 & 0 & 0 & 0 & 0 & 0 & 0 & -1 & i & i & 1 & i & 1 & -1 & i \\
 -1 & i & i & 1 & i & 1 & -1 & i & 0 & 0 & 0 & 0 & 0 & 0 & 0 & 0 \\
 0 & 0 & 0 & 0 & 0 & 0 & 0 & 0 & 0 & 0 & 0 & 0 & 0 & 0 & 0 & 0 \\
 0 & 0 & 0 & 0 & 0 & 0 & 0 & 0 & 0 & 0 & 0 & 0 & 0 & 0 & 0 & 0 \\
 0 & 0 & 0 & 0 & 0 & 0 & 0 & 0 & 0 & 0 & 0 & 0 & 0 & 0 & 0 & 0 \\
 0 & 0 & 0 & 0 & 0 & 0 & 0 & 0 & 0 & 0 & 0 & 0 & 0 & 0 & 0 & 0 \\
\end{array}
\right)
\end{eqnarray}
}
The 8 explicit  $4\times 4$ matrices are definitely tractable altough they have a rather complicated form that we do not write in explicit form.


\numberwithin{equation}{subsection}

\section{Short string limit for the folded string in $AdS_{5}\times S^{5}$}
\label{app:ads5-application}

This appendix is devoted to the calculation of $\mc O(\S)$ frequencies for the folded string in $AdS_{5}$.
As we mentioned in the main text, the aim of this application is to show that working at finite $\J>0$
with the correct fermionic fluctuation operator \cite{Iwashita:2010tg,Forini:2012bb} gives full agreement with the 
exact slope result derived by integrability methods \cite{Basso:2011rs, Gromov:2012eg,Gromov:2011bz}. The classical solution is derived in \cite{Frolov:2002av} and has the same form as in $AdS_{3}$.

\subsection{One-loop (quadratic) fluctuations}

Expanding the $AdS_5 \times S^5 $ superstring action in conformal gauge to quadratic order in the fluctuations near the folded spinning string
solution one finds  
\be
\widetilde{\mathscr S} = -\frac{\sqrt\lambda}{4\,\pi}\int d\tau\,\int_0^{2\,\pi} d\sigma\,(\widetilde {\cal L}_B + \widetilde {\cal L}_F),
\ee
where the fluctuation lagrangians are separately discussed in the next sections for bosons and fermions.

\subsubsection{Bosonic fluctuations}

The bosonic quadratic fluctuation Lagrangian is
\begin{eqnarray}
\widetilde {\cal L}_B &=& - \partial_a \td {t}\, \partial^a \td {t}- \mu_t^2\, \td {t}^2 +   \partial_a \td {\phi}\, \partial^a \td {\phi}+ \mu_{\phi}^2\, \td {\phi}^2\nonumber\\
&+& 4\, \td {\rho}\, (\kappa \sinh \rho\ \partial_0 \td {t} - w \cosh \rho\ \partial_0 \td {\phi})+  \partial_a \td {\rho} \,\partial^a \td {\rho}+\mu_{\rho}^2\, \td {\rho}^2\nonumber\\
&+& \partial_a {\beta}_u \,\partial^a {\beta}_u +\mu_{\beta}^2 \,{\beta}_u^2 +
 \partial_a {\varphi} \,\partial^a {\varphi}+\partial_a {\chi}_s \,\partial^a {\chi}_s + \J^2\, \chi_s^2 \ ,   
\end{eqnarray}
where
\ba
&&\mu_t^2= 2 \rho'^2 -\kappa^2 +\J^2,
 \qquad \mu^2_{\phi}=2 \rho'^2 -w^2 +\J^2,
 \qquad \mu^2_{\rho}=2 \rho'^2 -w^2-\kappa^2+2 \J^2, \nonumber \\
 && \mu_{\beta}^2=2 \rho'^2 +\J^2, \qquad \mu^{2}_{\chi_{s}}=\J^{2}.
 \ea
The two bosons $\beta_i$ ($i=1,2$) are two $AdS_5$ fluctuations transverse to the $AdS_3$ subspace in which the string is moving, while 
$\varphi,\chi_s$ ($s=1,2,3,4$) are five fluctuations in $S^5$.

We can write
\be
\wt {\cal L}_B = (\wt t, \wt \phi, \wt \rho)\,Q_{B}\,(\wt t, \wt \phi, \wt \rho)^T,
\ee
where the $Q_{B}$ operator is
\be
Q_{B} = \left(\begin{array}{ccc}
\partial^{2}-\mu_t^2 & 0 & 2\,\kappa\,\sinh \rho\,\partial_0 \\
0 & -\partial^{2}+\mu_\phi^2 & -2\,w\,\cosh\rho\,\partial_0 \\
-2\,\kappa\,\sinh \rho\,\partial_0 & 2\,w\,\cosh\rho\,\partial_0 &  -\partial^2+\mu_\rho^2
\end{array}
\right) .
\ee

\subsubsection{Fermionic fluctuations}

The fermionic lagrangian describes a system of 4+4 2d Majorana fermions. The result of \cite{Forini:2012bb}
reads
\be
\wt {\cal L}_F = 2\,i\,\overline\Psi\,D_{F}\,\Psi,\qquad D_{F} = \Gamma^{a}\,\partial_{a}+a(\sigma)\,\Gamma_{234}
+b(\sigma)\,\Gamma_{129},
\ee
where
\be
a(\sigma) = -\sqrt{\rho'^{2}+\J^{2}}, \qquad b(\sigma) = \frac{\J\,\kappa\,w}{2\,(\rho'^{2}+\J^{2})}.
\ee
Taking the square
\be
-D_{F}^{2} = -\partial^{a}\partial_{a}+a^{2}+b^{2}+\Gamma_{1234}\,a'+\Gamma_{29}(b'+2b\,\partial_{\sigma})
-2\,ab\,\Gamma_{1349}.
\ee
The matrices $\Gamma_{1234}$ and $\Gamma_{29}$ obey
\ba
&& \Gamma_{1234}^{2} = 1,\qquad \Gamma_{29}^{2} = -1,\qquad \Gamma_{1349}^{2}=1, \\
&& \{\Gamma_{1234},\Gamma_{29}\}=0, \qquad\{\Gamma_{1349},\Gamma_{29}\}=0, \qquad \{\Gamma_{1234}, \Gamma_{1349}\}=0,
\ea
and can be replaced (up to the tensor product with a multiple of the identity) by
\be
\Gamma_{1234} \equiv \sigma_{3},\qquad \Gamma_{29}\equiv i\,\sigma_{1}, \qquad \Gamma_{1349} = -\sigma_{2}.
\ee
In other words,
\be
Q_{F} = -D_{F}^{2}\equiv\left(\begin{array}{cc}
-\partial^{2}+a^{2}+b^{2}+a' & i\,(2ab+b'+2b\,\partial_{\sigma}) \\
i\,(-2ab+b'+2b\,\partial_{\sigma}) & -\partial^{2}+a^{2}+b^{2}-a'
\end{array}\right)
\ee

\subsection{The short string limit with $\J>0$}

We shall consider the short limit $\varepsilon\to 0$ with fixed $\J$. 
The solution of the equation of motion for $\rho(\sigma)$ is 
independent on $\J$ and reads
 \be
 \rho(\sigma) = \eps\,\sin\sigma+\frac{\eps^{3}}{12}\sin\sigma(\sin^{2}\sigma-3)+\frac{\eps^{5}}{320}
 \sin\sigma(4\sin^{4}\sigma-25\sin^{2}\sigma+45)+\dots.
 \ee
 It is convenient to collect the explicit expansions of various terms appearing in the fluctuation lagrangians.
 
 \subsubsection{Bosonic terms}
 
The expansion of the mixed terms in the operator $Q_{B}$ is
\ba
\kappa\,\sinh\rho &=& \eps\,\J\,\sin\sigma+\dots, \\
w\,\cosh\rho &=& \sqrt{\J^{2}+1}+\frac{\eps^{2}}{4}\,\frac{2(\J^{2}+1)\sin^{2}\sigma+1}{\sqrt{\J^{2}+1}}+\dots.
\ea
The expansion of the bosonic masses is
\ba
\mu_{t}^{2} &=& \eps^{2}(2\cos^{2}\sigma-1)+\dots, \\
\mu_{\phi}^{2} &=& -1+\eps^{2}\bigg(2\cos^{2}\sigma-\frac{1}{2}\bigg)+\dots,\\
\mu_{\rho}^{2} &=& -1+\eps^{2}\bigg(2\cos^{2}\sigma-\frac{3}{2}\bigg)+\dots,\\
\mu_{\beta}^{2} &=& \J^{2}+2\eps^{2}\cos^{2}\sigma+\dots.
\ea

It is convenient to rotate the $Q_{B}$ operator as 
\be
Q_{B} \to R^{-1}_{B}\,Q_{B}\,R_{B},\qquad R_{B} = \left(
\begin{array}{ccc}
 1 & 0 & 0 \\
 0 & \frac{1}{2} & \frac{1}{2} \\
 0 & -\frac{i}{2} & \frac{i}{2} \\
\end{array}
\right).
\ee
Then $Q_{B}=Q_{B}^{(0)}+\eps Q_{B}^{(1)}+\dots$ and 
\be
Q_{B}^{(0)} = \left(
\begin{array}{ccc}
 \omega ^2-n^2 & 0 & 0 \\
 0 & n^2-\omega ^2-2 \sqrt{\mathcal{J}^2+1} \omega -1 & 0 \\
 0 & 0 & n^2-\omega ^2+2 \sqrt{\mathcal{J}^2+1} \omega -1 \\
\end{array}
\right),
\ee
when acting on functions $\sim e^{in\sigma}$. This is 
a good starting point for perturbation theory. The important remark is that now it is possible to send 
$\J\to 0$. We find the same result by sending first $\J\to 0$ and then $\eps\to 0$. 
%
At finite $\J>0$, the eigenvalues of $Q_{B}^{(0)}$ can be written as  
\ba
\omega_{n} &=& \pm n, \\
\omega_{n} &=& \pm (\sqrt{n^{2}+\J^{2}}+\sqrt{1+\J^{2}}), \\
\omega_{n} &=& \pm (\sqrt{n^{2}+\J^{2}}-\sqrt{1+\J^{2}}) .
\ea

\subsubsection{Fermionic terms}

The expansion of the $a$ and $b$ functions appearing in the fermionic operator is
\ba
a(\sigma) &=& -\J-\frac{\eps^{2}\cos^{2}\sigma}{2\J}+\dots,\\
b(\sigma) &=& \frac{\sqrt{\J^{2}+1}}{2}+\eps^{2}\frac{\J^{2}-2(\J^{2}+1)\cos(2\sigma)}{8\J^{2}\sqrt{\J^{2}+1}}+\dots
\ea
Here we see a potential order of limits problem when $\eps, \J\to 0$ !

\bigskip

The explicit form of $Q_{F}^{(0)}$ is 
\be
Q_{F}^{(0)} =\left(
\begin{array}{cc}
 n^2+\frac{5 \mathcal{J}^2}{4}-\omega ^2+\frac{1}{4} & -(n+i \mathcal{J}) \sqrt{\mathcal{J}^2+1} \\
 i (i n+\mathcal{J}) \sqrt{\mathcal{J}^2+1} & n^2+\frac{5 \mathcal{J}^2}{4}-\omega ^2+\frac{1}{4} \\
\end{array}
\right).
\ee
Again, it is convenient to rotate $Q_{F}$ by a $n$-dependent rotation
\be
Q_{F}\to R^{-1}_{F}(n)\,Q_{F}\,R_{F}(n), \qquad R_{F}(n) = \left(
\begin{array}{cc}
 \sqrt{n^2+\mathcal{J}^2} & -\sqrt{n^2+\mathcal{J}^2} \\
 n-i \mathcal{J} & n-i \mathcal{J} \\
\end{array}
\right).
\ee
Then, acting on functions $\sim e^{in\sigma}$, the leading order is 
\be
Q_{F}^{(0)} = \left(
\begin{array}{cc}
 n^2+\frac{5 \mathcal{J}^2}{4}-\omega ^2-\sqrt{\left(\mathcal{J}^2+1\right) \left(n^2+\mathcal{J}^2\right)}+\frac{1}{4} & 0 \\
 0 & n^2-\omega ^2+\mathcal{J}^2 \left(\sqrt{\frac{n^2+\mathcal{J}^2}{\mathcal{J}^2+1}}+\frac{5}{4}\right)+\sqrt{\frac{n^2+\mathcal{J}^2}{\mathcal{J}^2+1}}+\frac{1}{4} \\
\end{array}
\right)
\ee
and its eigenfrequencies are 
\ba
\omega_{n} &=& \pm \bigg(\sqrt{n^{2}+\J^{2}}+\frac{1}{2}\sqrt{1+\J^{2}}\bigg), \\
\omega_{n} &=& \pm \bigg(\sqrt{n^{2}+\J^{2}}-\frac{1}{2}\sqrt{1+\J^{2}}\bigg) .
\ea

\subsection{Balance of the $\eps^{0}$ contributions}

In the flat space limit, we have the following contributions from the various fields (in the standard notation)
\be
\begin{array}{ccc}

\rm multiplicity & \rm field(s) & \omega_{n} \\ 
\hline\hline\\
1 & (t, \phi, \rho) & n \\
&& \sqrt{n^{2}+\J^{2}}+\sqrt{\J^{2}+1} \\
&& \sqrt{n^{2}+\J^{2}}-\sqrt{\J^{2}+1} \\ \\
2 & \beta_{u} & \sqrt{n^{2}+\J^{2}} \\ \\ \hline\\ 
1 & \varphi & n \\ 
4 & \chi_{s} & \sqrt{n^{2}+\J^{2}} \\ \\ \hline\hline\\
4 & \Psi &  \sqrt{n^{2}+\J^{2}}+\frac{1}{2}\,\sqrt{\J^{2}+1} \\
&& \sqrt{n^{2}+\J^{2}}-\frac{1}{2}\,\sqrt{\J^{2}+1} \\
2 & \rm ghost & n \\ \\ \hline\hline
\end{array}
\ee

\bigskip
Summing with weight $(-1)^{F}$ we find cancellation of (a)  massless contributions, (b) massive contributions 
$\sqrt{n^{2}+\J^{2}}$, (c) constant $n$-independent terms.

\subsection{Corrections to flat-space limit: Summary of results}
\label{sec:ads5-summary}

We can compute the correction to the eigenfrequencies 
\be
\omega_{n}(\eps) = \omega^{(0)}_{n}+\eps\, \omega^{(1)}_{n}+\eps^{2}\, \omega^{(2)}_{n}+\dots\,
\ee
by solving the (coupled) differential equations $\mathscr D(\partial_{\sigma}; \omega, n) \Phi_{n} = 0$
for the various field(s) $\Phi_{n}$. Imposing periodic boundary conditions we determing $\omega$, order
by order at small $\eps$. This procedure gives also the correction 
\be
\Phi_{n} = e^{i\,n\,\sigma}+\eps\,\Phi^{(1)}_{n}+\eps^{2}\,\Phi^{(2)}_{n}+\dots
\ee
We find a zero correction $\omega^{(1)}$ in all cases. The second order correction is smooth for  $|n|\neq 1$
otherwise some of the $\Phi^{(\ell)}_{n}$ coefficients can have singularities when $n\to \pm 1$.

For generic modes $|n|\neq 1$, the {\bf summary list} of the second order corrections is:
\ba
\omega_{n}^{(t, \rho, \phi), I} &=& n, \\
\label{eq:bose-II-correction}
\omega_{n}^{(t, \rho, \phi), II} &=& \sqrt{n^{2}+\J^{2}}+\sqrt{1+\J^{2}}+\eps^{2}\,\frac{2\J^{2}+2+\sqrt{(\J^{2}+1)(\J^{2}+n^{2})}}{4(\J^{2}+1)\sqrt{\J^{2}+n^{2}}}+\dots, \\
\label{eq:bose-III-correction}
\omega_{n}^{(t, \rho, \phi), III} &=& \sqrt{n^{2}+\J^{2}}-\sqrt{1+\J^{2}}+\eps^{2}\,\frac{2\J^{2}+2-\sqrt{(\J^{2}+1)(\J^{2}+n^{2})}}{4(\J^{2}+1)\sqrt{\J^{2}+n^{2}}}+\dots, \\
\omega_{n}^{\beta} &=& \sqrt{n^{2}+\J^{2}}+\eps^{2}\,\frac{1}{2\,\sqrt{\J^{2}+n^{2}}}+\dots, \\
\omega_{n}^{\varphi} &=& n, \\
\omega_{n}^{\chi} &=& \sqrt{\J^{2}+n^{2}}, \\
\label{eq:fermi-I-correction}
\omega_{n}^{\Psi, I} &=&  \sqrt{n^{2}+\J^{2}}+\frac{1}{2}\sqrt{1+\J^{2}}+\eps^{2}\,\frac{2\J^{2}+2+\sqrt{(\J^{2}+1)(\J^{2}+n^{2})}}{8(\J^{2}+1)\sqrt{\J^{2}+n^{2}}}+\dots, \\
\label{eq:fermi-II-correction}
\omega_{n}^{\Psi, II} &=&  \sqrt{n^{2}+\J^{2}}-\frac{1}{2}\sqrt{1+\J^{2}}+\eps^{2}\,\frac{2\J^{2}+2-\sqrt{(\J^{2}+1)(\J^{2}+n^{2})}}{8(\J^{2}+1)\sqrt{\J^{2}+n^{2}}}+\dots, \\
\omega_{n}^{\rm ghost} &=& n.
\ea

At the special values $n=\pm 1$ we find potential singularities in the corrections to $\Phi_{\pm 1}$
for various modes altough the correction $\omega^{(2)}_{\pm 1}$ is smooth. In all cases, with the exception of the bosonic mode $(t, \rho, \phi)^{III}$, what happens is that the true frequencies associated with $n=\pm 1$ obey 
\be
\omega_{n=1}^{(2), \rm true}+\omega_{n=-1}^{(2), \rm true} = 2\,\omega_{n=1}^{(2), \rm \ from\ the\ generic-n \ summary\ list}.
\ee
This means that we can safely use the summary list expressions if we are going to sum over all frequencies as is the case for the computation of the one-loop energy.

\bigskip
The only non trivial modification concerns $\omega_{n}^{(t, \rho, \phi), III}$ at $n=\pm 1$ that must be replaced
by zero. Indeed, for $|n|=1$ there are two independent periodic solutions that have precisely
$\omega^{(2)}\equiv 0$ as discussed in the following sections.
%

\bigskip
For $|n|\neq 1$ we have 
\be
\omega_{n}^{(t, \rho, \phi), I}+\omega_{n}^{(t, \rho, \phi), II}+\omega_{n}^{(t, \rho, \phi), III}+
2\,\omega_{n}^{\beta}+\omega_{n}^{\varphi}+4\,\omega_{n}^{\chi}-4\,\omega_{n}^{\Psi, I}-4\,\omega_{n}^{\Psi, II}-2\omega_{n}^{\rm ghost} =0,
\ee 
where the frequencies are those reported in the summary list.
Taking into account that $\kappa = \J + \dots$ and that $\eps^{2} = \frac{2\cS}{\sqrt{\J^{2}+1}}+\dots$, we find 
that the $|n|=1$ modes give
\ba
E_{1} &=& \frac{1}{2\kappa}\cdot 2\,\bigg(
\omega_{1}^{(t, \rho, \phi), I}+\omega_{1}^{(t, \rho, \phi), II}+
2\,\omega_{1}^{\beta}+\omega_{1}^{\varphi}+4\,\omega_{1}^{\chi}-4\,\omega_{1}^{\Psi, I}-4\,\omega_{1}^{\Psi, II}-2\omega_{1}^{\rm ghost}
\bigg)  = \nonumber \\
&=& -\frac{1}{\kappa}\,\omega_{1}^{(t, \rho, \phi), III} = 
-\frac{\cS}{2\,(\J+\J^{3})}+\mc O(\cS^{2}), 
\ea
in complete agreement with  \cite{Gromov:2011bz}.

\bigskip\bigskip
We conclude this long appendix by a detailed example of the frequency computation, including special low modes.

\subsection{A detailed example of calculation: The $\beta$ mode}
\subsubsection{Generic $n$}

This is the simplest case. One has to solve the equation 
\be
\bigg(\frac{d^{2}}{d\sigma^{2}}+\omega^{2}-\mu_{\beta}^{2}\bigg)\,\Phi_{n}(\sigma)=0,
\ee
where boundary conditions are periodic and the following perturbative Ansatz is imposed
\ba
\omega &=& \sqrt{n^{2}+\J^{2}}+\eps^{2}\omega^{(2)}_{n}+\dots, \\
\Phi_{n}(\sigma) &=& e^{i\,n\,\sigma}+\eps^{2}(z_{1}\,e^{i\,(n+2)\,\sigma}+z_{2}\,e^{i\,(n-2)\,\sigma})+\dots\, .
\ea
Solving the equation at the first non-trivial order determines for generic $n$
\be
z_{1} = -\frac{1}{8\,(n+1)},\qquad
z_{2} = \frac{1}{8\,(n-1)},\qquad
\omega_{n}^{(2)} = \frac{1}{2\,\sqrt{n^{2}+\J^{2}}}.
\ee

\subsubsection{Special values $n=\pm 1$}

One has to consider the special values $n=\pm 1$ at which the above solution breaks down. The problem is that 
for $n=\pm 1$ the solutions starting as $e^{\pm i\,\sigma}$ are mixed up. Thus, the correct Ansatz in this case is 
\ba
\omega &=& \sqrt{1+\J^{2}}+\eps^{2}\omega^{(2)}+\dots, \\
\Phi(\sigma) &=& e^{i\,\sigma}+\alpha\,e^{-i\,\sigma}+\eps^{2}(z_{1}\,e^{3\,i\,\sigma}+z_{2}\,e^{-3\,i\,\sigma})+\dots,
\ea
and also the mixing parameter $\alpha$ has to be determined. Plugging into the differential equation one finds two solutions
\ba
&& \alpha=+1, \quad z_{1}=+z_{2}=-\frac{1}{16},\quad \omega^{(2)} = \frac{3}{4\,\sqrt{1+\J^{2}}}, \\
&& \alpha=-1, \quad z_{1}=-z_{2}=-\frac{1}{16},\quad \omega^{(2)} = \frac{1}{4\,\sqrt{1+\J^{2}}}.
\ea
The sum of the two different values of $\omega^{(2)}$ is twice the naive value which is obtained by evaluating
the generic result $\omega^{(2)}_{n}$ at $n=1$. Hence, if we sum over all frequencies then we can simply
use the expression $\omega^{(2)}_{n}$.

\bigskip
Of course, in this simple case, the values $\alpha=\pm 1$ tell us that parity would have been enough to disentangle the two frequencies. Nevertheless, the above procedure is general.


\numberwithin{equation}{section}
\section{Separating out wrapping terms in infinite sums}
\label{app:sums}

We often have to compute complicated sums depending on $\J$ and we want to separate out the exponentially suppressed contribution at 
large $\J$. We illustrate how to practially treat sums of the kind occurring in the computation by discussing the nice example of 
\be
S(\mc J, \alpha) = \sum_{n=-\infty}^{\infty} f(n),\qquad f(n)=\frac{1}{(n^{2}-\alpha^{2})^{2}}\frac{1}{\sqrt{n^{2}+\J^{2}}},\qquad
0<\alpha<1, \ \J>0.
\ee
We would like to apply Euler-McLaurin formula since, up to a remainder that we don't write explicitly, 
\ba
&& \frac{1}{2}f(m)+f(m+1)+\dots+f(n-1)+\frac{1}{2}f(n) = \int_{m}^{n}f(x)dx \nonumber \\
&&\qquad +\sum_{k=2}^{\infty}
\frac{B_{k}}{k!}\bigg[
f^{(k-1)}(n)-f^{(k-1)}(m)
\bigg]+\mbox{remainder}.
\ea
When $m\to-\infty$ and $n\to +\infty$, the derivative terms vanish and only the integral remains. Actually, 
this means that there is a remainder in the Euler formula is generically nonzero, but is exponentially suppressed at large $\J$. However, the integral diverges at $n^{2}=\alpha^{2}$ and the singularity is not integrable. 

The trick is then to write 
\be
f(n) = \frac{1}{(n^{2}-\alpha^{2})}g(n^{2}), \qquad g(n^{2})=\frac{1}{\sqrt{n^{2}+\J^{2}}},
\ee
and to use
\ba
&&\sum_{n=-\infty}^{\infty}\frac{g(n^{2})}{(n^{2}-\alpha^{2})^{2}} = 
\sum_{n=-\infty}^{\infty}\frac{g(n^{2})-g(\alpha^{2})-(n^{2}-\alpha^{2})\,g'(\alpha^{2})}{(n^{2}-\alpha^{2})^{2}}
\nonumber \\
&& \qquad +g(\alpha^{2})\,\sum_{n=-\infty}^{\infty}\frac{1}{(n^{2}-\alpha^{2})^{2}} 
+g'(\alpha^{2})\,\sum_{n=-\infty}^{\infty}\frac{1}{n^{2}-\alpha^{2}} .
\ea
The last two terms can be summed exactly and the first sum can be converted into an integral up to the remainder.
The result is then
\ba
S(\J, \alpha) &=& \frac{\pi  \left(\cot (\pi  \alpha ) \left(2 \alpha ^2+\mathcal{J}^2\right)+\pi  \alpha  \csc ^2(\pi  \alpha ) \left(\alpha
   ^2+\mathcal{J}^2\right)\right)}{2 \alpha ^3 \left(\alpha ^2+\mathcal{J}^2\right)^{3/2}}\\
   &&+\frac{\frac{\left(2 \alpha ^2+\mathcal{J}^2\right) \left(\log \left(\frac{\sqrt{\alpha ^2+\mathcal{J}^2}}{\alpha }+1\right)-\log
   \left(\frac{\sqrt{\alpha ^2+\mathcal{J}^2}}{\alpha }-1\right)\right)}{2 \left(\alpha ^2+\mathcal{J}^2\right)^{3/2}}-\frac{\alpha
   }{\alpha ^2+\mathcal{J}^2}}{\alpha ^3}+\mbox{exponentially suppressed}\nonumber .
\ea


\section{Details of AC shifts of frequencies}
\label{app:AC-shifts}

The shifts appearing in the calculation of frequencies according to the quantization of the algebraic curve
are

\ba
\Delta\Omega_{S} &=& -\J, \\
\Delta\Omega_{1} &=& \left(-\sqrt{\mathcal{J}^2+1}-\mathcal{J}\right)+\left(-\frac{1}{2 \mathcal{J}^2+2}-\frac{1}{\mathcal{J} \sqrt{\mathcal{J}^2+1}}\right)
   \mathcal{S} \\
   && +\frac{\left(3 \mathcal{J}^5+9 \mathcal{J}^3+28 \sqrt{\mathcal{J}^2+1} \mathcal{J}^2+8 \sqrt{\mathcal{J}^2+1}+12
   \sqrt{\mathcal{J}^2+1} \mathcal{J}^4\right) \mathcal{S}^2}{16 \mathcal{J}^3
   \left(\mathcal{J}^2+1\right)^{5/2}}+O\left(\mathcal{S}^3\right), \nonumber \\
\Delta\Omega_{A} &=& \left(\sqrt{\mathcal{J}^2+1}-\mathcal{J}\right)+\left(\frac{1}{2 \mathcal{J}^2+2}-\frac{1}{\mathcal{J} \sqrt{\mathcal{J}^2+1}}\right)
   \mathcal{S} \\
   &&+\frac{\left(-3 \mathcal{J}^5-9 \mathcal{J}^3+28 \sqrt{\mathcal{J}^2+1} \mathcal{J}^2+8 \sqrt{\mathcal{J}^2+1}+12
   \sqrt{\mathcal{J}^2+1} \mathcal{J}^4\right) \mathcal{S}^2}{16 \mathcal{J}^3
   \left(\mathcal{J}^2+1\right)^{5/2}}+O\left(\mathcal{S}^3\right), \nonumber\\
\Delta\Omega_{3} &=& \left(\frac{\sqrt{\mathcal{J}^2+1}}{2}-\mathcal{J}\right)+\left(\frac{1}{4 \left(\mathcal{J}^2+1\right)}-\frac{1}{2 \mathcal{J}
   \sqrt{\mathcal{J}^2+1}}\right) \mathcal{S} \\
   &&+\frac{\left(-3 \mathcal{J}^5-9 \mathcal{J}^3+28 \sqrt{\mathcal{J}^2+1} \mathcal{J}^2+8
   \sqrt{\mathcal{J}^2+1}+12 \sqrt{\mathcal{J}^2+1} \mathcal{J}^4\right) \mathcal{S}^2}{32 \mathcal{J}^3
   \left(\mathcal{J}^2+1\right)^{5/2}}+O\left(\mathcal{S}^3\right), \nonumber \\
\Delta\Omega_{4} &=& \left(-\frac{\sqrt{\mathcal{J}^2+1}}{2}-\mathcal{J}\right)+\left(-\frac{1}{2 \mathcal{J} \sqrt{\mathcal{J}^2+1}}-\frac{1}{4
   \left(\mathcal{J}^2+1\right)}\right) \mathcal{S} \\
   && +\frac{\left(3 \mathcal{J}^5+9 \mathcal{J}^3+28 \sqrt{\mathcal{J}^2+1} \mathcal{J}^2+8
   \sqrt{\mathcal{J}^2+1}+12 \sqrt{\mathcal{J}^2+1} \mathcal{J}^4\right) \mathcal{S}^2}{32 \mathcal{J}^3
   \left(\mathcal{J}^2+1\right)^{5/2}}+O\left(\mathcal{S}^3\right). \nonumber 
\ea

\section{Details of AC-WS regularization matching}
\label{app:AC-details}

The quantities $\Delta^{(0)}_{A}$ and $\Omega^{(0)}_{A}$ defined in Section (\ref{sec:matching}) for the 
various polarizations $A=S, A, 1, 3, 4$ have the explicit values:
\ba
\Delta_{S}^{(0)} &=& 0, \\
\Delta_{A}^{(0)} &=& -\frac{\mathcal{S}}{\mathcal{J}^2}+\frac{\sqrt{\mathcal{J}^2+1} \mathcal{S}^2}{2 \mathcal{J}^4}+O\left(\mathcal{S}^3\right), \\
\Delta_{1}^{(0)} &=& \frac{\mathcal{S}}{\mathcal{J}^2}-\frac{\sqrt{\mathcal{J}^2+1} \mathcal{S}^2}{2 \mathcal{J}^4}+O\left(\mathcal{S}^3\right), \\
\Delta_{3}^{(0)} &=&-\frac{\mathcal{S}}{2 \mathcal{J}^2}+\frac{\sqrt{\mathcal{J}^2+1} \mathcal{S}^2}{4 \mathcal{J}^4}+O\left(\mathcal{S}^3\right), \\
\Delta_{4}^{(0)} &=& \frac{\mathcal{S}}{2 \mathcal{J}^2}-\frac{\sqrt{\mathcal{J}^2+1} \mathcal{S}^2}{4 \mathcal{J}^4}+O\left(\mathcal{S}^3\right),
\ea
and
\ba
\Omega^{(0)}_{S} &=& -\mathcal{J}+O\left(\mathcal{S}^3\right), \\
\Omega^{(0)}_{A} &=& -\mathcal{J}-\frac{\left(\frac{\mathcal{J}}{\sqrt{\mathcal{J}^2+1}}+1\right) \mathcal{S}}{\mathcal{J}^2} \\
&& +\frac{\left(3 \mathcal{J}^6+8
   \mathcal{J}^4+7 \mathcal{J}^2+2 \sqrt{\mathcal{J}^2+1} \mathcal{J}+3 \sqrt{\mathcal{J}^2+1} \mathcal{J}^5+7 \sqrt{\mathcal{J}^2+1}
   \mathcal{J}^3+2\right) \mathcal{S}^2}{4 \mathcal{J}^4 \left(\mathcal{J}^2+1\right)^{5/2}}+O\left(\mathcal{S}^3\right), \nonumber\\
\Omega^{(0)}_{1} &=& -\mathcal{J}+\frac{\left(1-\frac{\mathcal{J}}{\sqrt{\mathcal{J}^2+1}}\right) \mathcal{S}}{\mathcal{J}^2}\\
&& +\frac{\left(-3 \mathcal{J}^6-8
   \mathcal{J}^4-7 \mathcal{J}^2+2 \sqrt{\mathcal{J}^2+1} \mathcal{J}+3 \sqrt{\mathcal{J}^2+1} \mathcal{J}^5+7 \sqrt{\mathcal{J}^2+1}
   \mathcal{J}^3-2\right) \mathcal{S}^2}{4 \mathcal{J}^4 \left(\mathcal{J}^2+1\right)^{5/2}}+O\left(\mathcal{S}^3\right),\nonumber \\
\Omega^{(0)}_{3} &=& -\mathcal{J}-\frac{\left(\frac{\mathcal{J}}{\sqrt{\mathcal{J}^2+1}}+1\right) \mathcal{S}}{2 \mathcal{J}^2}\\
&& +\frac{\left(3 \mathcal{J}^6+8
   \mathcal{J}^4+7 \mathcal{J}^2+2 \sqrt{\mathcal{J}^2+1} \mathcal{J}+3 \sqrt{\mathcal{J}^2+1} \mathcal{J}^5+7 \sqrt{\mathcal{J}^2+1}
   \mathcal{J}^3+2\right) \mathcal{S}^2}{8 \mathcal{J}^4 \left(\mathcal{J}^2+1\right)^{5/2}}+O\left(\mathcal{S}^3\right),\nonumber \\
\Omega^{(0)}_{4} &=& -\mathcal{J}+\frac{\left(1-\frac{\mathcal{J}}{\sqrt{\mathcal{J}^2+1}}\right) \mathcal{S}}{2 \mathcal{J}^2}\\
&& +\frac{\left(-3 \mathcal{J}^6-8
   \mathcal{J}^4-7 \mathcal{J}^2+2 \sqrt{\mathcal{J}^2+1} \mathcal{J}+3 \sqrt{\mathcal{J}^2+1} \mathcal{J}^5+7 \sqrt{\mathcal{J}^2+1}
   \mathcal{J}^3-2\right) \mathcal{S}^2}{8 \mathcal{J}^4 \left(\mathcal{J}^2+1\right)^{5/2}}+O\left(\mathcal{S}^3\right).\nonumber
\ea

\bibliography{AC-Biblio}{}
\bibliographystyle{JHEP}

\end{document}